\documentclass[a4paper,fleqn]{cas-sc}

\usepackage{siunitx}
\usepackage[acronym,toc]{glossaries}
\usepackage[authoryear]{natbib}
\usepackage{subcaption}

\newacronym{TRL}{TRL}{technology readiness level}
\newacronym{TCO}{TCO}{total cost of ownership}
\newacronym{PnP}{PnP}{plug-and-play}

\newacronym{DC}{DC}{direct current}
\newacronym{LVDC}{LVDC}{low-voltage DC}
\newacronym{MVDC}{MVDC}{medium-voltage DC}
\newacronym{AC}{AC}{alternating current}
\newacronym{LVAC}{LVAC}{low-voltage AC}
\newacronym{MVAC}{MVAC}{medium-voltage AC}
\newacronym{HFAC}{HFAC}{high-frequency AC}
\newacronym{QoS}{QoS}{quality of supply}

\newacronym{UC}{UC}{ultra-capacitor}
\newacronym{ESS}{ESS}{energy storage system}
\newacronym{BESS}{BESS}{battery energy storage system}
\newacronym{HESS}{HESS}{hybrid energy storage system}
\newacronym{EOL}{EOL}{end-of-life}
\newacronym{BOL}{BOL}{begin-of-life}
\newacronym{SoC}{SoC}{state-of-charge}
\newacronym{DoD}{DoD}{depth of discharge}
\newacronym{FC}{FC}{fuel cell}
\newacronym{PEMFC}{PEMFC}{proton-exchange membrane fuel cell}
\newacronym{PV}{PV}{photovoltaic}
\newacronym{DG}{DG}{diesel generator}
\newacronym{ICE}{ICE}{internal combustion engine}
\newacronym{GT}{GT}{gas turbine}
\newacronym{CPL}{CPL}{constant power load}
\newacronym{PPL}{PPL}{pulsed power load}
\newacronym{CI}{CI}{cold ironing}
\newacronym{UPS}{UPS}{uninterruptible power supply}
\newacronym{RES}{RES}{renewable energy source}

\newacronym[firstplural=shipboard power systems (SPS)]{SPS}{SPS}{shipboard power system}
\newacronym[firstplural=zero emission ships (ZES)]{ZES}{ZES}{zero emission ship}
\newacronym[firstplural=integrated power systems (IPS)]{IPS}{IPS}{integrated power system}
\newacronym[firstplural=all-electric ships (AES)]{AES}{AES}{all-electric ship}
\newacronym{DP}{DP}{dynamic positioning}
\newacronym{HEV}{HEV}{hybrid electric vehicle}

\newacronym[firstplural=energy management strategies (EMS)]{EMS}{EMS}{energy management strategy}
\newacronym{PSC}{PSC}{power system control}
\newacronym{PI}{PI}{proportional-integral}
\newacronym{MPC}{MPC}{model predictive control}
\newacronym{ECMS}{ECMS}{equivalent consumption minimization strategy}
\newacronym{SDP}{SDP}{sequential dynamic programming}
\newacronym{SVM}{SVM}{space-vector modulation}
\newacronym{PWM}{PWM}{pulse-width modulation}
\newacronym{FOC}{FOC}{field-oriented control}
\newacronym{SMC}{SMC}{sliding mode control}
\newacronym{MAS}{MAS}{multi-agent system}
\newacronym{DSM}{DSM}{demand side management}
\newacronym{MPPT}{MPPT}{maximum power point tracking}
\newacronym{LPF}{LPF}{low-pass filter}
\newacronym{HPF}{HPF}{high-pass filter}

\newacronym{PSO}{PSO}{particle swarm optimization}

\newacronym{HiL}{HiL}{hardware-in-the-loop}
\newacronym{RCP}{RCP}{rapid control prototyping}

\newacronym{IMO}{IMO}{International Maritime Organization}
\glsdisablehyper

\begin{document}
\let\WriteBookmarks\relax
\def\floatpagepagefraction{1}
\def\textpagefraction{.001}
\shorttitle{Degradation-aware Predictive EMS for Fuel Cell-Battery Ship Power Systems}
\shortauthors{T. Kopka et~al.}

\title [mode = title]{Degradation-aware Predictive Energy Management for Fuel Cell-Battery Ship Power System with Data-driven Load Forecasting}                      


\author[1]{Timon Kopka}
\cormark[1]
\ead{t.kopka@tudelft.nl}

\author[1]{Sara Tamburello}

\author[2]{Luca Oneto}

\author[1]{Lindert {van Biert}}

\author[1]{Henk Polinder}

\author[1]{Andrea Coraddu}

\affiliation[1]{organization={Department of Maritime and Transport Technology, Delft University of Technology},
	addressline={Mekelweg 2}, 
	city={2628 CD Delft},
	country={The Netherlands}}

\affiliation[2]{organization={Department of Informatics, Bioengineering, Robotics and Systems Engineering, Università degli Studi di Genova},
	addressline={Via Dodecaneso 35}, 
	city={16146 Genova},
	country={Italy}}

\cortext[cor1]{Corresponding author: t.kopka@tudelft.nl}

\begin{abstract}
	Hydrogen-based zero-emission ships are a key element in the decarbonization of the maritime sector.
	To strengthen these systems' economic competitiveness, it is key to drive their costs to a minimum.
	Current literature mainly focuses on fuel consumption minimization, but there is a lack of explicit consideration of costs arising from cell degradation and optimization-based approaches that leverage information on future load trajectories.
	This work aims at minimizing the operational cost of fuel cell(FC)-battery hybrid shipboard power systems, accounting for hydrogen consumption and cell degradation as the main cost drivers. 
	A degradation-aware predictive energy management strategy (EMS) utilizing data-driven load forecasting is designed and showcased at the example of a virtually retrofitted harbor tug.
	This work shows that the real onboard measurements of the vessel can be utilized to make accurate load predictions over 15min.
	Results indicate that the degradation-aware, predictive control simultaneously reduces the hydrogen consumption by up to \SI{5.8}{\%} and the cell degradation by up to \SI{36.4}{\%} with an aged FC system when compared to a filter-based benchmark applied to real operating data of the harbor tug.
	With an increased prediction horizon of \SI{1}{h}, further significant reductions of \SI{3.8}{\%} and \SI{14.0}{\%} could be shown.
\end{abstract}




\maketitle

\section{Introduction}
Electrification and hybridization are two key enablers for increasing the operational flexibility and capability of onboard energy systems in ships~\citet{nuchtureeEnergyEfficiencyIntegrated2020}.
\Glspl{FC} are a prominent candidate for the replacement of conventional diesel engines as main generators, aiming at the decarbonization of maritime transport~\citet{shakeriHydrogenFuelCells2020}.
\Glspl{ESS} allow for the decoupling of the load and generation, making it possible to operate the main supplies more steadily and efficiently, while high power gradients and power peaks can be compensated by, e.g., batteries~\citet{mutarrafEnergyStorageSystems2018}.
This can be especially beneficial in ship types whose load profile is highly volatile, changing quickly within a short time frame, e.g. harbor tugs.

One fundamental line of research for hybrid vessels is the coordination of the generation and storage systems, as the batteries introduce additional degrees-of-freedom for the control system. 
The main challenge for the \gls{EMS} is the optimal dispatch of power sources over time, ensuring that each component- and system-level constraints are respected~\citet{geertsmaDesignControlHybrid2017}.
Typically, the \gls{EMS} aims at the minimization of the fuel consumption, while ensuring that the load power can always be supplied by the generation side~\citet{jaurolaOptimisingDesignPower2019}.
However, with the introduction of \glspl{FC}, an additional key performance criterion is the cell degradation, which is required to be minimized such that their lifetime is extended, keeping stack replacement costs low~\citet{shakeriHydrogenFuelCells2020,wallnofer-ogrisMainDegradationMechanisms2024}.
Hence, the optimal energy management of \gls{FC}-battery \gls{SPS} poses a multi-objective optimization problem.

In the literature, a multitude of approaches attempts to solve this problem. 
Generally, methods can be divided into heuristic and optimization-based approaches~\citet{xieOptimizationBasedPowerEnergy2022}.
Many studies on the energy management for \gls{FC}-battery based \gls{SPS} focus on PI or filter-based controllers~\citet{balestraEnergyManagementStrategies2021,balestraModellingSimulationZeroemission2021,kopkaVirtualImpedancebasedFrequency2025} to achieve peak-shaving for steady \gls{FC} operation or on efficiency improvement via rule-based controllers~\citet{hanEnergyManagementSystem2014,bassamImprovedEnergyManagement2016}.
Similarly, fuzzy-logic controllers are utilized to translate expert insights into rules for a functioning \gls{EMS}~\citet{zhuFuzzyLogicBased2014,zhaoImprovedFuzzyLogic2022}.

More advanced \glspl{EMS} are optimization-based, which can either be instantaneous or predictive, where the latter attempts to include knowledge of future system states and disturbances, e.g. the load power, in the solution of the energy management problem.
An example of instantaneous optimization is \gls{ECMS}, which conventionally is applied to hybridized diesel-based power systems~\citet{sciarrettaControlHybridElectric2007,grimmeliusControlHybridShip2011,yuanEquivalentConsumptionMinimization2016}, but has also been shown to be applicable to zero-emission ships~\citet{zhangRealTimeOptimizationEnergy2020,kopkaHierarchicalControlStrategy2023}.
\citet{xiangECMSMultiObjectiveEnergy2021} shows the implementation of \gls{ECMS} for multi-objective optimization.
The advantage of instantaneous optimization is that no knowledge of future system disturbances is required.
However, in case a prediction of the future mission profile is possible, it can be incorporated in the \gls{EMS} to improve the controller performance and explicitly include system limitations.
An example of this is the use of \gls{MPC}, which can be utilized in various functions, such as optimal control of parallel prime movers~\citet{paranMPCbasedPowerManagement2015}, and optimal power split for hybrid systems~\citet{haseltalabModelPredictiveManeuvering2019,khazaeiOptimalFlowMVDC2021}.
\citet{xieMPCinformedECMSBased2021} and \citet{antonopoulosMPCFrameworkEnergy2021} formulate \gls{ECMS} implementations whose parameters are set by overarching predictive or mission-scale optimizers.
\gls{MPC} is commonly implemented in multiple layers, decomposing the optimization problem into varying timescales.
\citet{haseltalabModelPredictiveManeuvering2019,zhangTwolevelModelPredictive2022} decouple the optimal steady-state dispatch from the dynamic power sharing in two-level \gls{MPC} formulations whereas a three-level implementation is proposed by~\citet{seenumaniHierarchicalOptimalControl2010}.
Predictive energy management for zero-emission ships is much less investigated. In \citet{banaeiComparativeAnalysisOptimal2020,banaeiCostEffectiveOperation2020} an \gls{MPC} is applied to determine the steady-state economic dispatch for a zero-emission ferry.
The work in \citet{banaeiStochasticModelPredictive2021} further incorporates \gls{FC} degradation as an objective.
However, the mentioned applications for \gls{FC}-battery systems apply the optimization for large time steps while assuming a mission-scale quantification of load uncertainties. This disregards both the highly fluctuating load dynamics, as well as the optimal power split in dynamic conditions and uncertainties between predicted and actual load.

Most studies dedicated to the optimized control of hydrogen-based \gls{SPS} only optimize for fuel consumption.
Besides the cost of hydrogen, the cell degradation and eventual stack replacement of the \glspl{FC}, as well as increased consumption due to degradation, are key drivers of the overall operating costs of a hydrogen \gls{FC}-based system.
An explicit quantification of the cell degradation and its minimization during dynamic operation is done by~\citet{fletcherEnergyManagementStrategy2016}. Similarly, \citet{bassamDevelopmentMultischemeEnergy2017} analyzes the cell degradation for various rule-based \gls{EMS} in an \gls{SPS}. However, these studies lack an attempt to quantify the costs arising from the cell degradation and, furthermore, do not include their degradation function in a predictive \gls{EMS}.

The implementation of an \gls{MPC} requires an estimate of future states, inputs and/or disturbances.
Many sources assume perfect knowledge of future load to showcase the effectiveness of predictive control.
Typically, this approach leads to promising results. 
However, in practice, knowledge of future system states and disturbances is limited.
Accordingly, real systems require designers of a predictive \gls{EMS} to develop methods for predicting the load, which can be based on heuristics, arise from the mission planning, or be based on operational data.
A promising approach is the utilization of data-driven techniques for predicting the load profile over a certain time interval into the future, denoted as the prediction horizon.
This challenge can be characterized as a Time Series Forecasting (TSF) problem~\citet{lim2021time,shalev2014understanding,bishop2023deep}.
Many approaches exist for solving TSF~\citet{lim2021time,shalev2014understanding,bishop2023deep}.
Classical methods such as Autoregressive Integrated Moving Average (ARIMA) and Exponential Smoothing often yield good results~\citet{hyndman2018forecasting}, but Machine Learning (ML) has become an increasingly attractive option due to its potential to improve model accuracy and scale to larger datasets with higher dimensionalities.
ML models can be broadly categorized into two main families.
The first is Shallow ML, which can achieve reasonably good performance with small to medium-sized datasets, provided that domain-specific features are carefully engineered~\citet{shalev2014understanding}.
The second family, Deep ML, can automatically learn features, i.e., representations without the need for manual feature engineering.
However, leveraging Deep ML models can be challenging due to the large number of architectural design choices and the substantial amount of data required~\citet{bishop2023deep}.
Recent studies have demonstrated that ML-based approaches, particularly those using Deep ML, can outperform classical methods on a variety of TSF tasks~\citet{oreshkin2019n,salinas2020deepar}.

Unfortunately, Deep ML models rely on massive labeled datasets, and constructing such datasets is challenging due to limited data availability (i.e., data may not be available at all) or high data annotation costs (i.e., data must be labeled by humans).
To address these challenges, Foundation Models (FMs), such as Large Language Models in Natural Language Processing~\citet{zhao2023survey} and advanced models in Computer Vision~\citet{awais2023foundational}, have emerged as powerful paradigms capable of achieving state-of-the-art performance in their respective fields.
The success of these FMs stems from their ability to leverage vast amounts of data to learn general-purpose representations, enabling them to be fine-tuned or even used directly in a zero-shot manner across a diverse range of downstream tasks~\citet{gruver2024large}.
This approach not only reduces the need for task-specific model development but also encapsulates a broad understanding of the world, endowing these models with exceptional versatility and efficiency.
Motivated by this success, researchers have started exploring the development of Time Series Foundation Models (TSFMs) to harness the potential of the foundation model paradigm for time series data.
By capitalizing on large-scale time series datasets, TSFMs aim to achieve superior performance across a variety of TSF tasks, offering a unified framework that can accelerate research and application development in this field~\citet{liang2024foundation}.
The no-free-lunch theorem~\citet{adam2019no} asserts that identifying the optimal algorithm for a specific application requires evaluating multiple approaches.

This work formulates an \gls{MPC} that handles second-scale load fluctuations and continuously prescribes optimized power references for \glspl{FC} and batteries considering their dynamic and steady-state constraints and operational costs.

In our work, we explored several state-of-the-art methods and, based on their performance, proposed a novel approach that combines their strengths.

In our work, we propose the following contributions to address the gaps which have been identified in the literature and the state-of-the art for predictive energy management of hydrogen-based \gls{SPS}:
\begin{itemize}
	\item First, a degradation-aware optimization for a \gls{FC}-battery hybrid \gls{SPS} is constructed and applied in an ECMS-based \gls{EMS} in real-time. The optimization minimizes operational costs arising from hydrogen consumption and cell degradation as main cost drivers.
	
	\item Second, several state-of-the art TSF methods are explored for the load forecasting of a harbor tug, utilizing the system's available real-time measurements. Based on their performance, a novel approach that combines their strengths is proposed to provide accurate load predictions.
	
	\item Third, the data-driven load forecasting and degradation-aware cost function are formulated in an \gls{MPC} which explicitly minimizes the operating costs over the prediction horizon.
\end{itemize}

The remainder of this manuscript is organized as follows.
Chapter~\ref{sec:sys-descr} introduces the original layout and measurement of the harbor tug and describes the virtual retrofit for obtaining an all-electric \gls{FC}-battery hybrid vessel which serves as the case study in this work.
Dynamic component and system models as well as a derivation of a complete state-space model are presented in chapter~\ref{sec:modeling}. Further, this chapter introduces cost functions for hydrogen consumption and cell degradation and introduces system constraints.
Chapter~\ref{sec:load-forecast} describes the development of the data-driven load forecasting method and showcases its performance in predicting the future load of the harbor tug.
The optimization function for the degradation-aware energy management is derived in chapter~\ref{sec:power-sys-ctrl}. Both an \gls{ECMS}-based and a model predictive \gls{EMS} as well as a filter-based benchmark controller are described here.
Numerical results of mission simulations for the display of performance using the proposed strategies are analyzed  and discussed in chapter~\ref{sec:num-investigations}.
Finally, chapter~\ref{sec:conclusions} summarizes the main conclusions of this work.

\section{System description}
\label{sec:sys-descr}
The \gls{EMS} developed for this study is applicable to fully electric \gls{SPS} with \glspl{FC} and \gls{ESS} on the generation side. Here, a harbor tug is considered, whose operating profile is characterized by a highly volatile load, with distinct operating modes and short mission profiles.
As tugs typically operate in fleets and stay within the harbor space, they are a viable candidate for \gls{FC}-battery systems, assuming that a hydrogen bunkering infrastructure is present.
This chapter first describes the original diesel-hybrid harbor tug and the available measured data from its operation.
In the next step, based on original design and mission requirements, virtual retrofit of the tugboat's power system with \glspl{FC} and batteries is described, yielding the vessel regarded in this work.

\subsection{Original System}
The original harbor tug is equipped with a hybrid propulsion architecture and hybrid energy system with its specifications listed in Table~\ref{tab:original-specs}. The main diesel engines are sized to carry the bulk of the propulsive load while the shaft motors and batteries allow for load leveling, gradient support, and full-electric operation at low load as well as a power take-off for feeding electric consumers.

\begin{figure}
	\centering
	\includegraphics{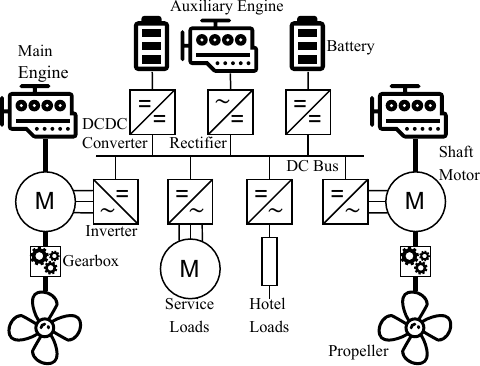}
	\caption{Primary systems of original diesel-based SPS}
	\label{fig:original-topology}
\end{figure}

Twelve months of operational data has been analyzed to determine the required power of the considered vessel for broad variety of actual mission profiles. Over this period, a total number of $167$ missions have been logged with a logging rate of \SI{5}{s}. Decisive for the power system control, as outlined in Section~\ref{sec:power-sys-ctrl}, is the time-dependent total power load. However, a broader array of real-time measurements is available and required for an accurate data-driven load forecasting. More detailed descriptions of the available data-sets and their utilization for the load forecasting are given in Section~\ref{sec:load-forecast}.

\begin{table}
	\centering
	\caption{Specifications of original diesel-hybrid power system.}
	{\begin{tabular}{lr}
			\hline
			Parameter & Info \\
			\hline
			Type & Harbor tug \\
			Length & \SI{29.0}{m} \\
			Width & \SI{10.5}{m} \\
			Displacement & \SI{575}{t} \\
			Max. Speed & \SI{13.5}{kn} \\
			Propulsion & Diesel-hybrid \\
			Main Engine & 2x\SI{1840}{kW} \\
			Shaft Motor & 2x\SI{115}{kW} \\ 
			Auxiliary Engine & \SI{800}{kW} \\
			Battery Pack & 2x\SI{120}{kWh} \\
			\hline
	\end{tabular}}
	\label{tab:original-specs}
\end{table}

\begin{figure}
	\centering
	\includegraphics{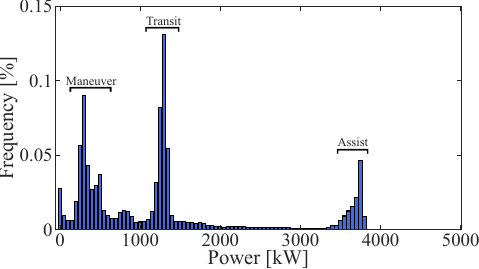}
	\caption{Histogram of total load measured during operation of the harbor tug}
	\label{fig:data-histogram}
\end{figure}

\begin{figure}
	\centering
	\includegraphics{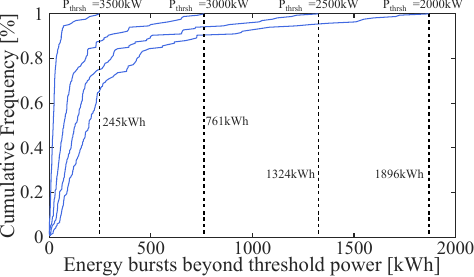}
	\caption{Cumulative frequency of energy bursts beyond specified threshold power during assist}
	\label{fig:energy-burst}
\end{figure}

\subsection{System Retrofit}
\label{subsec:retrofit}
\begin{figure}
	\centering
	\includegraphics{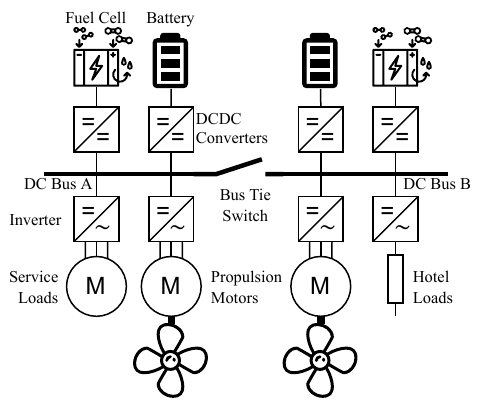}
	\caption{Primary systems of retrofitted full electric FC-hybrid power system}
	\label{fig:sps-topology}
\end{figure}

Based on the original power system and the operational requirements, a hydrogen \gls{FC}-based re-design has been conceptualized.
Two interconnected DC busses, each attached to an \gls{FC} and a battery system comprises the complete energy generation system. All loads are electrified, and fed by power converters.
The resulting topology is displayed in Fig.\ref{fig:sps-topology} and the specifications listed in Table~\ref{tab:retrofit-specs}.

Regarding the sizing of components, the total \gls{FC} power at \gls{EOL} shall be sufficient for providing the maximum required power during assist.
The maximum measured power, safe for negligibly short peaks, is \SI{3800}{kW}, covering both the propulsive and auxiliary power load.
Assuming that the \gls{EOL} power of the \gls{FC} systems is at \SI{90}{\%} of the system's original power rating~\citet{kleenDOEHydrogenProgram2024}, two proton-exchange membrane \gls{FC} (PEMFC) systems, each with a maximum power output of \SI{2125}{kW} at \gls{BOL}, are selected. This leaves a total of \SI{3825}{kW} available generative power before a stack replacement.

\gls{FC} systems show increased degradation at high power operation beyond \SI{80}{\%} of their maximum rating. Hence, it is desirable to size the energy storage such that the \glspl{FC} are rarely forced above this threshold.
At \gls{EOL}, this amounts to \SI{3060}{kW} of total power from the \glspl{FC}.
Bursts of energy that require power beyond a certain threshold $P_{thresh}$ are shown in Fig.~\ref{fig:energy-burst}. For the desired operation limit of the \glspl{FC}, an additional energy of up to \SI{761}{kWh} is required to cover all observed mission profiles. With an assumed depth-of-discharge of \SI{60}{\%}, this requires \SI{1268}{kWh} of installed energy storage capacity. For the \gls{ESS} sizing, we round this value to \SI{1250}{kWh}, distributed over two units. To achieve sufficient capability for compensating short-term power bursts, and power gradient support, a charge and discharge rating of \SI{1.5}{C} is sufficient. Here, we select lithium iron phosphate (LFP) batteries due to their high efficiency and stable degradation behavior.

\begin{table}
	\caption{Specifications of re-designed full-electric FC-battery power system.}
	\begin{tabular}{lr}
		\hline
		Parameter & Info \\
		\hline
		Type & Harbor tug \\
		Propulsion & Full-electric \\
		Fuel Cells & 2x\SI{2125}{kW} (BOL) \\
		Battery Pack & 2x\SI{625}{kWh} (BOL) \\
		Propulsion Motors & 2x\SI{1840}{kW} \\
		\hline
		\centering
	\end{tabular}
	\label{tab:retrofit-specs}
\end{table}

\section{Power System Model}
\label{sec:modeling}
This section describes the dynamic models used for the primary components of the \gls{SPS} which encompasses the \gls{FC}, the \gls{ESS}, electric load and the power distribution system.
From these models, a state-space representation of the complete system is derived, to be used for the optimization-based formulation of the system's energy management.
In addition, approximations of the hydrogen consumption and the cell degradation of the \glspl{FC}, and a cost model for aggregating both factors in the operating costs are described.

\begin{figure}
	\centering
	\includegraphics{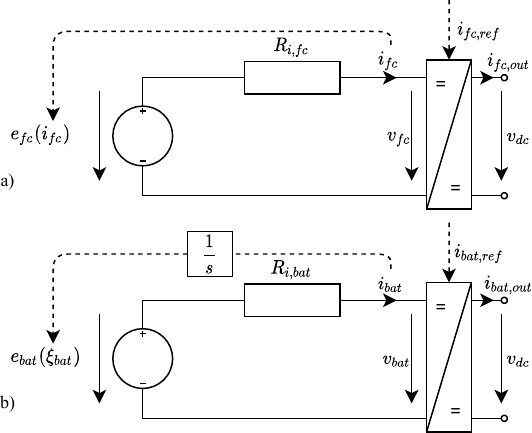}
	\caption{Simulation models for the PEMFC (a) and battery (b) including current-controlled power electronics converter}
	\label{fig:sim-models}
\end{figure}

\begin{table}
	\centering
	\caption{Fuel Cell and Battery Parameters}
	\begin{center}
		\begin{tabular}{ccc}
			\hline
			Parameter & Description & Value \\
			\hline
			$P_{fc,max}$ & Max. net FC power & \SI{4150}{kW} \\
			$\dot{P}_{fc,max}$ & Max. FC power gradient& \SI{212.5}{kW/s} \\
			$R_{i,fc}$ & FC resistance (BOL/EOL) & \SI{23.2}{}/\SI{28.0}{m\Omega} \\
			\hline
			$C_{bat}$ & Battery Capacity & \SI{3125}{Ah} \\
			$R_{i,bat}$ & Bat. internal resistance & \SI{2.4}{m\Omega} \\
			$I_{bat,max}$ & Maximum current & \SI{9.4}{kA} \\
			\hline
		\end{tabular}
		\label{tab:fc-bat-params}
	\end{center}
\end{table}

\subsection{Fuel Cell}
\begin{figure}
	\centering
	\includegraphics{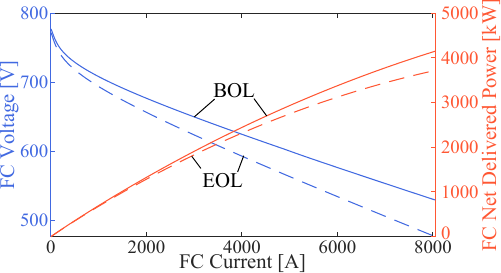}
	\caption{PEMFC voltage $v_{fc}$ and net output power as functions of output current $i_{fc}$ for BOL and EOL conditions}
	\label{fig:fc_polarization_curve}
\end{figure}

In this work PEM\glspl{FC} serve as main power generation devices. 
The equivalent circuit of the simulation model consisting of an equivalent voltage source and internal resistance is shown in Fig.~\ref{fig:sim-models} a), as proposed by~\citet{njoyaGenericFuelCell2009} for dynamic simulation of \gls{FC}-based systems.
The voltage source $e_{fc}$ is current-dependent and computed as
\begin{equation}
	e_{fc}(i_{fc})=E_{oc,fc}-N_{s,fc}E_{ts}ln(\frac{i_{fc}}{N_{s,fc}I_0})\cdot\frac{1}{sT_d/3+1}
\end{equation}
where stack settling time $T_d$, Tafel slope $E_{ts}$, exchange current $I_0$, number of series-connected cells $N_{s,fc}$, and open circuit voltage $E_{oc,fc}$, as well as the internal resistance $R_{i,fc}$, accounting for the linear voltage decrease, are extracted from the manufacturer datasheet~\citet{nedstackNedstackFCS13XXL2023} according to the method described by~\citet{njoyaGenericFuelCell2009}.
With this parameterization, the output voltage $v_{fc}$ follows the polarization curve of the PEM\gls{FC} module as described in its datasheet. Table~\ref{tab:fc-bat-params} lists the \gls{FC} model parameters. The voltage and power curves over output current for the system at \gls{BOL} and \gls{EOL} are shown in Fig.~\ref{fig:fc_polarization_curve}.

\subsection{Energy Storage System}
\begin{figure}
	\centering
	\includegraphics{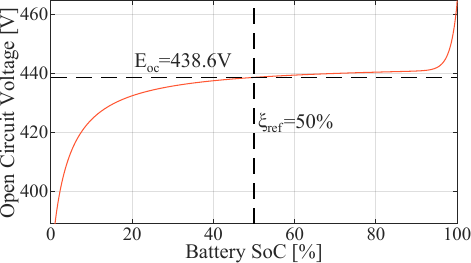}
	\caption{Open-circuit voltage of LFP battery system}
	\label{fig:sim-bat-pol}
\end{figure}

An LFP battery system is integrated into the power system as \gls{ESS}. 
The simulation model, proposed by~\citet{tremblayGenericBatteryModel2007}, is similar to the \gls{FC} model and is shown in Fig.~\ref{fig:sim-models} b). Here, the equivalent voltage source $e_{bat}$ is depending on the battery \gls{SoC}, which is calculated via Coulomb counting.
The system is parameterized based on the manufacturer datasheet~\citet{rs-proRechargeableLithiumionIron} and its open-circuit voltage is displayed in Fig.~\ref{fig:sim-bat-pol}.
The battery model parameters are shown in Table~\ref{tab:fc-bat-params}.
The LFP technology is characterized by its flat \gls{SoC}-dependency curve. This enables the approximation of the open-circuit voltage with a constant value $E_{oc}=e_{bat}(\xi_{bat}=50\%)$.

\subsection{State Space Modeling}
\begin{figure}
	\centering
	\includegraphics{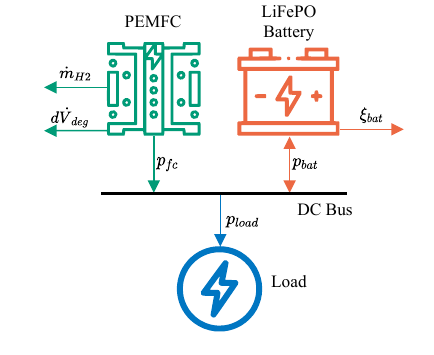}
	\caption{Simplified one-node control model for the EMS}
	\label{fig:control-model}
\end{figure}

For the energy management, the control model of the \gls{SPS} is simplified as a single-node system connected to a PEM\gls{FC}, a battery, and an aggregated electric load, as shown in Fig.~\ref{fig:control-model}.
The \gls{FC} model generates an estimate of hydrogen consumption $\dot{m}_{H2}$ and cell voltage degradation $\Delta V_{deg}$, as described in the modeling section~\ref{sec:modeling}.
Both values are taken into account as objectives to be minimized in the energy management.

Furthermore, the PEM\gls{FC} and the LiFePO battery feed power into the bus, denoted as $p_{fc}$ and $p_{bat}$, respectively.
We define the \gls{FC} power $p_{fc}$ and the battery \gls{SoC} $\xi_{bat}$ as system states.
Our control input is the \gls{FC} power gradient $\dot{p}_{fc}$.
Whereas $p_{fc}$ must always be $\geq 0$, $p_{bat}$ may be negative, indicating a charging of the battery. Furthermore, a limit is set on the input, i.e., $\dot{p}_{fc}$. Hence, system inputs and states must adhere to the following constraints
\begin{equation}
	|\dot{p}_{fc}(n)| \leq \dot{P}_{fc,max}
	\label{eq:Pfc_grad_lim}
\end{equation}
\begin{equation}
	0 \leq p_{fc}(n) \leq P_{fc,max}
	\label{eq:Pfc_lim}
\end{equation}
\begin{equation}
	SoC_{min} \leq \xi(n) \leq SoC_{max} \quad \forall \quad n \in [k, k+N]
	\label{eq:soc_lim}
\end{equation}
The \gls{FC} power dynamics are simply computed as
\begin{equation}
	p_{fc}(n+1)=p_{fc}(n)+\dot{p}_{fc}(n)T_s
\end{equation}
where $T_s$ is the discretization interval.
The power generation must always be in balance with the load power $p_{load}$. The latter is modeled as an external disturbance and since $p_{fc}$ is defined as a system state, we can compute the battery power and current as
\begin{equation}
	p_{bat}(n)=p_{load}(n) - p_{fc}(n) - \frac{T_s}{2}\dot{p}_{fc}(n)
	\label{eq:Pbat}
\end{equation}
\begin{equation}
	i_{bat}(n)=\frac{E_{oc}}{2R_{i,bat}}-\sqrt{-\frac{p_{bat}}{R_{i,bat}}+(\frac{E_{oc}}{2R_{i,bat}})^2}
	\label{eq:Ibat}
\end{equation}
\begin{equation}
	\xi_{bat}(n+1)=\xi_{bat}(n)-\frac{i_{bat}(n)}{C_{bat}}T_s
\end{equation}
for $i_{bat}R_{i,bat} << E_{oc}$ we can observe that $v_{bat}\simeq E_{oc}$ and simplify~(\ref{eq:Ibat}) with a linearized function for the battery current
\begin{equation}
	i_{bat}(n)\simeq \frac{p_{bat}}{E_{oc}} = \frac{p_{load}(n) - p_{fc}(n) - \frac{T_s}{2}\dot{p}_{fc}(n)}{E_{oc}}
\end{equation}
We can rewrite the complete system dynamics as
\begin{equation}
	[p_{fc}(n+1),\ \xi_{bat}(n+1)]^T = A[p_{fc}(n),\ \xi_{bat}(n)]^T + B\dot{p}_{fc}(n) + B_d p_{load}(n)
	\label{eq:sys-dynamics}
\end{equation}
\begin{align}
	A&=
	\begin{bmatrix}
		1 & 0\\
		\frac{-T_s}{E_{oc}C_{bat}} & 1\\
	\end{bmatrix}
	&
	B&=
	\begin{bmatrix}
		T_s\\
		\frac{-T_s}{2E_{oc}C_{bat}}\\
	\end{bmatrix}
	&
	B_d&=
	\begin{bmatrix}
		0\\
		\frac{T_s}{E_{oc}C_{bat}}\\
	\end{bmatrix}
\end{align}

\subsection{Hydrogen Consumption}
\label{sec:modeling-h2-consumption}
The hydrogen consumption of the stack is directly connected to the \gls{FC} current $I_{fc,in}$, as laid out in~\citet{kopkaHierarchicalControlStrategy2023}.
In addition, hydrogen is lost due to hydrogen crossover and internal short-circuits in the cells~\citet{maoInvestigationPolymerElectrolyte2017}.
On a system level, the \gls{FC} consumes auxiliary power for its balance-of-plant, which depends on the operation of the system~\citet{shakeriHydrogenFuelCells2020}. Since it is relevant for the power balance, this work regards the \gls{FC} power $p_{fc}$ as the net delivered power, which is the power delivered by the stack minus the auxiliary power.

\begin{figure}
	\centering
	\includegraphics{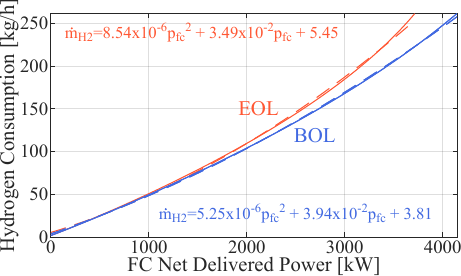}
	\caption{Hydrogen consumption and quadratic approximation for FC system at BOL and EOL}
	\label{fig:hydrogen-consumption}
\end{figure}

For the optimization-based power system control, the hydrogen consumption can be approximated with a second-order polynomial as a function of net delivered power $\dot{m}_{H2}(p_{fc})$.
The modeled hydrogen consumption and its quadratic approximations at \gls{BOL} and \gls{EOL} are shown in Fig.~\ref{fig:hydrogen-consumption}. The curves show the effect of degradation on the increased hydrogen need, which becomes especially apparent in the high power region.

\subsection{FC degradation}
\label{sec:modeling-fc-degradation}
\begin{figure}
	\centering
	\includegraphics{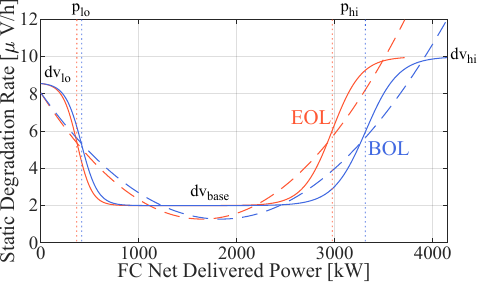}
	\caption{Cell-level static voltage degradation of FC system (solid) and approximation with second-order polynomial (dashed) at BOL and EOL}
	\label{fig:static-degradation}
\end{figure}

Apart from the hydrogen consumption, a key performance indicator for the control is the \gls{FC} degradation, affecting the state-of-health and lifetime of the PEM\gls{FC}.
PEM\gls{FC} degradation is a complex combination of mechanisms, affected by several causes, which ultimately results into a decrease of the delivered output power. 
In this work, degradation is evaluated in terms of voltage decrease over time at a rated current.
On an operational level, the load is considered the main cause affecting PEM\gls{FC} degradation. 
To distinguish its effects on the PEM\gls{FC} system, the load is typically broken down into three main categories, i.e., degradation arising from static operation, dynamic load changes, and system start-up cycles.
In literature, different degradation rates are reported for operation in low and high current operation regions, and a considerably lower static degradation in between those regions.
Representative degradation values and region limits are listed in Table~\ref{tab:fc-deg-params}.
The change of degradation rate between the regions is approximated with a linear function around the region limit.
To make the static degradation function easier to handle in an optimization, it is approximated via a second-order polynomial, enforcing its convexity over the power range.
The static degradation of the \gls{FC} over its operation range and its quadratic optimization at \gls{BOL} and \gls{EOL} are shown in Fig.~\ref{fig:static-degradation}.
Since the power thresholds are relative to the maximum power, the curve is compressed towards the origin with increased degradation.

\begin{table}
	\centering
	\caption{FC Degradation Parameters (per cell)}
	\begin{center}
		\begin{tabular}{cccc}
			\hline
			Param. & Description & Value & Ref. \\
			\hline
			$p_{lo}$ & Low power threshold & \SI{20}{\%} & \citet{shakeriHydrogenFuelCells2020}\\
			$p_{hi}$ & High power threshold & \SI{80}{\%} & \citet{fletcherEnergyManagementStrategy2016}\\
			$dv_{hi}$ & High power deg. (>$p_{hi}$) & \SI{10.0}{\mu V/h} & \citet{chenLifetimePredictionEconomic2015}\\
			$dv_{lo}$ & Low power deg. (<$p_{lo}$)& \SI{8.6}{\mu V/h} & \citet{chenLifetimePredictionEconomic2015}\\
			$dv_{base}$ & Base degradation & \SI{2.0}{\mu V/h} & \citet{debruijnReviewDurabilityDegradation2008}\\
			$dv_{dyn}$ & Dynamic degradation & $0.5\cdot$\SI{19}{\mu V} & \citet{mengDynamicCurrentCycles2021}\\
			$dt_{dyn}$ & Reference ramp time & \SI{10}{s} & (assumed)\\
			\hline
		\end{tabular}
		\label{tab:fc-deg-params}
	\end{center}
\end{table}

The dynamic degradation is a result of load changes. \citet{mengDynamicCurrentCycles2021} reports an average of $\SI{19}{\mu V}$ per complete power cycle from idling to full power and back.
No studies were made on the effect of the current gradient in dynamic degradation studies. However, we assume a quadratic relationship between the gradient and the degrading effect, meaning the faster the change in output current, the higher the degrading effect of the dynamic change.
For this study, we assume a reference time of \SI{10}{s} at which the reported degradation rate is realized. Additionally, the dynamic degradation is scaled linearly for partial current cycles:

\begin{equation}
	d\dot{V}_{dyn}(\dot{p}_{fc})=\frac{dt_{dyn}dv_{dyn}}{P_{fc,max}^2}\dot{p}_{fc}^2:=dv_{dyn2}\dot{p}_{fc}^2
	\label{eq:dynamic-deg}
\end{equation}

Degradation due to start-up and shutdown cycles are disregarded in this study, since it is assumed that the \gls{FC} is not switched during the operation, therefore not affecting the control scheme.

\subsection{Cost Model}
\label{sec:modeling-cost-model}
Hydrogen consumption and \gls{FC} degradation are two objectives for the operation of the hybrid ship. It is essential to determine a suitable trade-off between these two for the design of the power system control. In this study, the related costs for both objectives are estimated to enable a comparison between the two.
The cost estimate for hydrogen $c_{deg}$ is straightforward since the fuel is a traded commodity.

\begin{figure}
	\centering
	\includegraphics{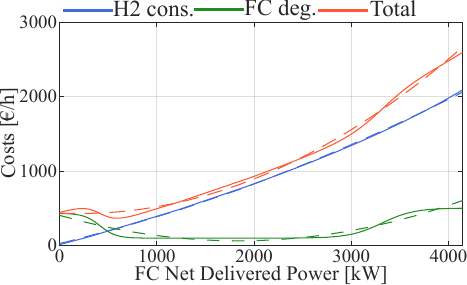}
	\caption{Cost curves of hydrogen consumption and static degradation. Model data (solid) and approximations with second-order polynomials (dashed) at BOL.}
	\label{fig:cost_estimate}
\end{figure}

The cost for degradation is more difficult to define. This study estimates it as the cost of reduced lifetime due to voltage degradation.
Conventionally, a power loss of \SI{10}{\%} at rated current compared to a new system is proposed as the end-of-life of \gls{FC} systems in heavy duty applications with a targeted lifetime of \SI{25000}{h}~\citet{kleenDOEHydrogenProgram2024}. This is challenged by~\citet{jouinEstimatingEndoflifePEM2016} as being too restrictive and, in dynamic applications, reached considerably earlier.
Due to a lack of alternative standards, we use the limit of \SI{10}{\%} for this study.
Accordingly, at reaching \SI{90}{\%} of its original maximum power rating, the \gls{FC} stack needs to be replaced. The degradation-specific cost is estimated as the partial replacement costs due to operation-specific voltage degradation:
\begin{equation}
	c_{deg}=\frac{r_{fc,stack}c_{fc,capex}P_{fc,max}N_s}{r_{eol}V_{fc,min}}
\end{equation}
A comparison of hydrogen costs, static degradation costs and total costs from static operation of the \gls{FC} system is displayed in Fig.~\ref{fig:cost_estimate}.
The cost parameters used in this study are listed in Table~\ref{tab:cost-params}.

\begin{table}
	\centering
	\caption{Cost Parameters for hydrogen consumption and cell degradation}
	\begin{center}
		\begin{tabular}{ccc}
			\hline
			Parameter & Description & Value \\
			\hline
			$c_{h2}$ & Hydrogen cost & \SI{8}{\text{€}/kg} \\
			$c_{fc,capex}$ & FC system cost & \SI{1500}{\text{€}/kW} \\
			$r_{fc,stack}$ & FC stack cost ratio & \SI{50}{\%} \\
			$r_{eol}$ & FC power loss at EOL  & \SI{10}{\%} \\
			$c_{deg}$ & Degradation cost  & \SI{50.6}{\text{€}/\mu V} \\
			\hline
		\end{tabular}
		\label{tab:cost-params}
	\end{center}
\end{table}

\section{Load Forecasting}
\label{sec:load-forecast}
This section deals with the problem of predicting the load power of the vessel for the future with a sampling frequency of \SI{5}{s}, matching the data-logging rate of the original measurements.
In particular, Section~\ref{subsec:loadforecasting:data} will describe in detail the problem and the data available to address it while Section~\ref{subsec:loadforecasting:methods} will explain the proposed methods based on advanced ML techniques to solve the problem.
\subsection{Problem Description and Available Data}
\label{subsec:loadforecasting:data}
The case study of this paper examines a vessel featuring a hybrid propulsion system that includes diesel and electric power.
The proposed methodology utilizes operational data logged by the vessel's automation system.
The data under consideration comprises $5$ million samples with a sampling rate of $5$ seconds, spanning over one year of operational data, between 24th of April 2022 and 6th of May 2023.
It consists of $18$ input features, as presented in Table~\ref{tab:load:data}, where the feature total power ($\mathrm{p_{tot}}$) is defined as the sum of power, i.e., the total power to be forecasted is obtained as
\begin{equation}
\mathrm{p_{tot}} = (\gamma_{de,s}+\gamma_{de,p})P_{de,max} + \gamma_{aux}P_{aux,max} + \mathrm{p_{bat,s}} + \mathrm{p_{bat,p}}.
\end{equation}

\begin{table}
\setlength{\tabcolsep}{0.15cm}
\renewcommand{\arraystretch}{1.2}
\centering
\caption{Load forecasting: available data on the original vessel.}
\label{tab:load:data}
\begin{tabular}{clcc}
\hline
Symbol & Definition & Unit \\
\hline
$\mathrm{v}$ & Vessel Speed & [$\mathrm{knots}$] \\
$\mathrm{n_{prop,s}}$ & Propeller Speed - Starboard & [$\mathrm{rpm}$] \\
$\mathrm{n_{prop,p}}$ & Propeller Speed - Port & [$\mathrm{rpm}$] \\
$\mathrm{n_{em,s}}$ & Electric motor speed - Starboard & [$\mathrm{rpm}$] \\
$\mathrm{n_{em,p}}$ & Electric motor speed - Port & [$\mathrm{rpm}$] \\
$\mathrm{p_{em,s}}$ & Electric motor power - Starboard & [$\mathrm{kW}$] \\
$\mathrm{p_{em,p}}$ & Electric motor power - Port & [$\mathrm{kW}$] \\
$\mathrm{p_{bat,s}}$ & Battery power - Starboard & [$\mathrm{kW}$] \\
$\mathrm{p_{bat,p}}$ & Battery power - Port & [$\mathrm{kW}$] \\
$\mathrm{\xi_{s}}$ & State of charge - Starboard & [\%] \\
$\mathrm{\xi_{p}}$ & State of charge - Port & [\%] \\
$\mathrm{n_{de,s}}$ & Diesel engine speed - Starboard & [$\mathrm{rpm}$] \\
$\mathrm{n_{de,p}}$ & Diesel engine speed - Port & [$\mathrm{rpm}$] \\
$\mathrm{n_{aux}}$ & Auxiliary generator set speed & [$\mathrm{rpm}$] \\
$\mathrm{\dot{m}_{de,s}}$ & Diesel fuel consumption - Starboard & [$\mathrm{kg/min}$] \\
$\mathrm{\dot{m}_{de,p}}$ & Diesel fuel consumption - Port & [$\mathrm{kg/min}$] \\
$\mathrm{\dot{m}_{aux}}$ & Auxiliary generator set fuel consumption & [$\mathrm{kg/min}$] \\
$\mathrm{\gamma_{de,s}}$ & Diesel engine load - Starboard & [\%] \\
$\mathrm{\gamma_{de,p}}$ & Diesel engine load - Port & [\%]  \\
$\mathrm{\gamma_{aux}}$ & Auxiliary generator set load & [\%] \\
$\mathrm{\delta_{s}}$ & Steering angle - Starboard & [$^{\circ}$] \\
$\mathrm{\delta_{p}}$ & Steering angle - Port & [$^{\circ}$] \\
$\mathrm{p_{tot}}$ & Total Power & [$\mathrm{kW}$] \\
\hline
\end{tabular}
\end{table}

In this work, the vessel has been retrofitted as described in Section~\ref{subsec:retrofit}, and, as a result, only the information provided in Table~\ref{tab:load:data_retrofit} is available. 
Specifically, we assume that the retrofit to a fully electric configuration does not alter the vessel’s operational profile. 
Consequently, parameters such as $\mathrm{v}$, $\mathrm{\delta_{p}}$, $\mathrm{\delta_{s}}$, and $\mathrm{p_{tot}}$ are assumed to remain unchanged.

\begin{table}
\setlength{\tabcolsep}{0.15cm}
\renewcommand{\arraystretch}{1.2}
\centering
\caption{Load forecasting: available data after the vessel retrofitting.}
\label{tab:load:data_retrofit}
\begin{tabular}{clcc}
\hline
Symbol & Definition & Unit \\
\hline
$\mathrm{v}$ & Vessel Speed & [$\mathrm{knots}$] \\
$\mathrm{n_{prop,s}}$ & Propeller speed - Starboard & [$\mathrm{rpm}$] \\
$\mathrm{n_{prop,p}}$ & Propeller speed - Port & [$\mathrm{rpm}$] \\
$\mathrm{\delta_{s}}$ & Steering angle - Starboard & [$^{\circ}$] \\
$\mathrm{\delta_{p}}$ & Steering angle - Port & [$^{\circ}$] \\
$\mathrm{p_{tot}}$ & Total Power & [$\mathrm{kW}$] \\
\hline
\end{tabular}
\end{table}

As a consequence, after the retrofitting, our scope is to predict, at time $t_0$, the $\mathrm{p_{tot}}$ for the next $\Delta^+$ minutes with a sampling frequency of \SI{5}{s}, i.e., $\mathrm{p_{tot}}(t)$ with $t \in \{t_0 + 5s, t_0 + 10s, \cdots, t_0 + \Delta^+ \}$, based on the information of $\mathrm{v}$, $\mathrm{n_{prop,s}}$, $\mathrm{n_{prop,p}}$, $\mathrm{\delta_{p}}$, $\mathrm{\delta_{s}}$, and $\mathrm{p_{tot}}$ with $t \in [t_0 - \Delta^-, t_0]$ and $\Delta^-$ (see Section~\ref{subsec:loadforecasting:methods}).
Table~\ref{tab:load:data_retrofit_forecasting} summarizes inputs and output of our TSF problem.
\begin{table}
\setlength{\tabcolsep}{0.15cm}
\renewcommand{\arraystretch}{1.2}
\centering
\caption{Load forecasting: input and output of the forecast problem available at time $t_0$.}
\label{tab:load:data_retrofit_forecasting}
\begin{tabular}{cccccc}
\hline
Input/Output & Symbol & Frequency & Time Span \\
\hline
\multirow{5}{*}{Input}
& $\mathrm{v}$ & \multirow{5}{*}{\SI{5}{s}} & \multirow{5}{*}{$[t_0 - \Delta^-, t_0]$} \\
& $\mathrm{n_{prop,s}}$ && \\
& $\mathrm{n_{prop,p}}$ && \\
& $\mathrm{\delta_{s}}$ && \\
& $\mathrm{\delta_{p}}$ && \\
\hline
Output
& $\mathrm{p_{tot}}$ & \SI{5}{s} & $[t_0, t_0 + \Delta^+]$ \\
\hline
\end{tabular}
\end{table}
\subsection{Methods}
\label{subsec:loadforecasting:methods}
\begin{table*}
\setlength{\tabcolsep}{0.2cm}
\renewcommand{\arraystretch}{1.4}
\centering
\footnotesize
\caption{Load Forecasting: MAE, MAPE, and PPMCC for XGBoost, TCN, TSFM, and XGBoost+TSFM for different time horizons.}
\label{tab:exp:forecasting:1}

\begin{tabular}{cc}
    \hline
    \textbf{MAE}&
    \begin{tabular}{>{\centering\arraybackslash}p{2.5cm}  
		                    >{\centering\arraybackslash}p{2cm} 
		                    >{\centering\arraybackslash}p{2cm} 
		                    >{\centering\arraybackslash}p{2cm} 
		                    >{\centering\arraybackslash}p{2cm}}
	    \textbf{Model} & \textbf{5s} & \textbf{1min} & \textbf{15min} & \textbf{30min} \\ 
	    \hline
	    XGBoost       & $0.23 {\pm} 0.02$ & $0.23 {\pm} 0.02$ & $2.31 {\pm} 0.25$ & $28.74 {\pm} 3.26$ \\ 
	    TCN           & $0.22 {\pm} 0.02$ & $0.25 {\pm} 0.03$ & $2.35 {\pm} 0.14$ & $30.74 {\pm} 2.76$ \\ 
	    TSFM          & $0.41 {\pm} 0.04$ & $0.43 {\pm} 0.05$ & $2.85 {\pm} 0.23$ & $51.94 {\pm} 5.01$ \\ 
	    XGBoost+TSFM  & $0.17 {\pm} 0.02$ & $0.18 {\pm} 0.02$ & $1.19 {\pm} 0.06$ & $21.64 {\pm} 2.36$ \\ 
	    \end{tabular}\\
    \hline
    \textbf{MAPE}&
    \begin{tabular}{>{\centering\arraybackslash}p{2.5cm}  
		                    >{\centering\arraybackslash}p{2cm}  
		                    >{\centering\arraybackslash}p{2cm}  
		                    >{\centering\arraybackslash}p{2cm}  
		                    >{\centering\arraybackslash}p{2cm}}
	    XGBoost       & $1.2 {\pm} 0.1$ & $1.6 {\pm} 0.1$ & $4.4 {\pm} 0.5$ & $58.0 {\pm} 5.9$ \\ 
	    TCN           & $1.2 {\pm} 0.1$ & $1.7 {\pm} 0.2$ & $4.4 {\pm} 0.5$ & $61.0 {\pm} 7.1$ \\ 
	    TSFM          & $10.1 {\pm} 0.6$ & $11.0 {\pm} 1.4$ & $37.0 {\pm} 3.2$ & $97.0 {\pm} 7.4$ \\ 
	    XGBoost+TSFM  & $0.4 {\pm} 0.1$ & $0.4 {\pm} 0.1$ & $2.8 {\pm} 0.3$ & $51.0 {\pm} 5.9$ \\ 
	    \end{tabular}\\
    \hline
    \textbf{PPMCC}&
    \begin{tabular}{>{\centering\arraybackslash}p{2.5cm}  
		                >{\centering\arraybackslash}p{2cm}  
		                >{\centering\arraybackslash}p{2cm}  
		                >{\centering\arraybackslash}p{2cm}  
		                >{\centering\arraybackslash}p{2cm}}
	    XGBoost       & $0.998 {\pm} 0.012$ & $0.997 {\pm} 0.011$ & $0.926 {\pm} 0.011$ & $0.127 {\pm} 0.001$ \\ 
	    TCN           & $0.999 {\pm} 0.008$ & $0.998 {\pm} 0.009$ & $0.928 {\pm} 0.010$ & $0.129 {\pm} 0.001$ \\ 
	    TSFM          & $0.986 {\pm} 0.009$ & $0.985 {\pm} 0.007$ & $0.877 {\pm} 0.009$ & $0.099 {\pm} 0.001$ \\ 
	    XGBoost+TSFM  & $0.997 {\pm} 0.009$ & $0.998 {\pm} 0.013$ & $0.938 {\pm} 0.008$ & $0.139 {\pm} 0.001$ \\ 
	    \end{tabular}\\                             
    \hline
\end{tabular}
\end{table*}

The problem in Section~\ref{subsec:loadforecasting:data} can be effectively framed as a classical TSF problem~\citet{lim2021time,shalev2014understanding,bishop2023deep}.
TSF involves predicting future values of a variable of interest at time $t_0$, specifically for $t \in [t_0, t_0 + \Delta^+]$.
The forecasting approach depends on the available information at time $t_0$.
Predictions can be made based solely on the past values of the target variable (i.e., without covariates).
Alternatively, forecasts can leverage both past values of the target variable and past and/or future values of related variables (i.e., with covariates).
Including relevant covariates can improve forecast accuracy by providing context and capturing variations not explained by the target variable alone.
In some scenarios, forecasts can be generated using only covariates if past values of the target variable are unavailable.
In our specific case, as shown in Table~\ref{tab:load:data_retrofit_forecasting}, we have access to both past values of the target variable and past covariates.

\color{black}
As a first step, we tested a TSF method based on top-performing algorithms from different ML paradigms: Shallow ML (i.e., XGBoost~\citet{chen2016xgboost}), Deep ML (i.e., Temporal Convolutional Networks - TCN~\citet{bai2018empirical}), and TSFMs (i.e., the TSFM proposed in~\citet{dasdecoder}).
The first two approaches were implemented using the Dart library~\citet{herzen2022darts}, while the last one was implemented using the library released by~\citet{dasdecoder}.
Each of the algorithms we have selected and listed is characterized by a set of hyperparameters that require tuning.
\begin{itemize}
	\item XGBoost~\citet{chen2016xgboost} requires tuning the learning rate $l_r$, the maximum tree depth $n_d$, the minimum loss reduction $m_l$, the number of training samples per tree $n_b$, and the number of features $n_f$ randomly sampled at each node.
	\item TCN is an implementation of a dilated temporal convolutional network for forecasting, inspired by~\citet{bai2018empirical}.
	In this network, we tune the number of hidden layers $n_t$, the number of filters $f_i$ and the kernel size $k_i$ in each hidden layer, with $i \in \{ 1, \dots, n_h \}$, as well as the dropout rate $d_{r,i}$ of each hidden layer.
	The model is trained using the ADAM optimizer~\citet{kingma2014adam}, tuning the learning rate $l_r$ and the number of epochs $n_e$.
	\item TSFM~\citet{dasdecoder} is a pretrained time-series foundation model developed by Google Research for time-series forecasting.
	We fine-tune it using the ADAM optimizer~\citet{kingma2014adam}, tuning the learning rate $l_r$ and the number of epochs $n_e$.
\end{itemize}
The summary of these hyperparameters with the associated search space is reported in Table~\ref{table:hyper}.
\begin{table}
	\footnotesize
	\setlength{\tabcolsep}{0.2cm}
	\renewcommand{\arraystretch}{1.4}
	\centering
	\caption{\color{black}Hyperparameters and hyperparameters search space for all algorithms tested in this work.}
	\label{table:hyper}
	\begin{tabular}{|l|l|l|}
		\hline
		\hline
		\textbf{Algorithm} & \textbf{Hyperparameters} \\
		\hline
		\hline
		\multirow{3}{*}{XGBoost} 
		& $l_r: \{0.1, 0.05, 0.01, 0.005, 0.001\}$,\\
		& $n_d: \{3,5,10\}$, 
		$m_l: \{0,0.1,0.2\}$, \\
		& $n_b: \{0.6 n , 0.8 n, 1 n \}$,
		$n_f: \{0.5 d, 0.8 d, 1 d\}$ \\
		\hline
		\multirow{3}{*}{TCN}
		& $n_t: \{ 1, 2, 4, 8\}$, $f_i: \{ 4, 8, 16, 32\}$, $k_i: \{ 3, 5, 7\}$,\\
		& $d_{r,i}: \{0.001, 0.002, 0.01, 0.02, 0.1, 0.2 \}$, \\
		& $l_r: \{0.1, 0.05, 0.01, 0.005, 0.001\}$, $n_e: \{1, \cdots, 10^4 \}$ \\
		\hline
		\multirow{1}{*}{TSFM}
		& $l_r: \{0.1, 0.05, 0.01, 0.005, 0.001\}$, $n_e: \{1, \cdots, 10^4 \}$ \\
		\hline
		\hline
		- & $\Delta^-: \{ 5,10,30,60,350,600,1800 \}$ sec \\
		\hline
		\hline
	\end{tabular}
\end{table}

Algorithms hyperparameter tuning (tougher with $\Delta^-$) and performance evaluation, i.e., model selection and error estimation~\citet{OnetoB003}, were conducted using the standard temporal split of the train (60\%), validation (20\%), and test (20\%) sets commonly applied in TSF~\citet{herzen2022darts,hyndman2018forecasting,lim2021time}.
The training set is used to train the model, while the validation set is used to identify the optimal combination of hyperparameters.
Then, the training and validation sets are combined to train a final model using the best hyperparameter configuration, and its accuracy is evaluated on the test set, which is a fresh set of future data that has remained untouched throughout the process.
The accuracy is evaluated using a combination of quantitative and qualitative metrics.
The quantitative metrics include the Mean Absolute Error (MAE), the Mean Absolute Percentage Error (MAPE), and the Pearson Product-Moment Correlation Coefficient (PPMCC)~\citet{naser2021error}. 
The qualitative metrics comprise a scatter plot comparing actual versus predicted values~\citet{sainani2016value}, as well as an analysis of the real-time series behavior against the forecast generated by the best-performing model.

From the results of these models (Section~\ref{subsec:exp:loadforecasting}), we observe that TSF methods based on XGBoost and TCN outperformed the TSFM-based approach, with XGBoost slightly surpassing TCN.
Nevertheless, given the specific nature of our TSF problem, TSFM showed impressive performance.
Consequently, we propose using the TSFM-based forecast as a future covariate for the XGBoost-based TSF method.
This integration (see Section~\ref{subsec:exp:loadforecasting}) leads to a notable performance improvement in the XGBoost-based TSF, confirming the potential of TSFM in real-world TSF applications.
As a final remark, note that we set $\Delta^+ = \SI{15}{min}$ as, after that, the performance started to become too low to be used in practice.
\color{black}
\subsection{Application to Case Study}
\label{subsec:exp:loadforecasting}
\begin{figure}
\centering
\includegraphics[width=.4\textwidth]{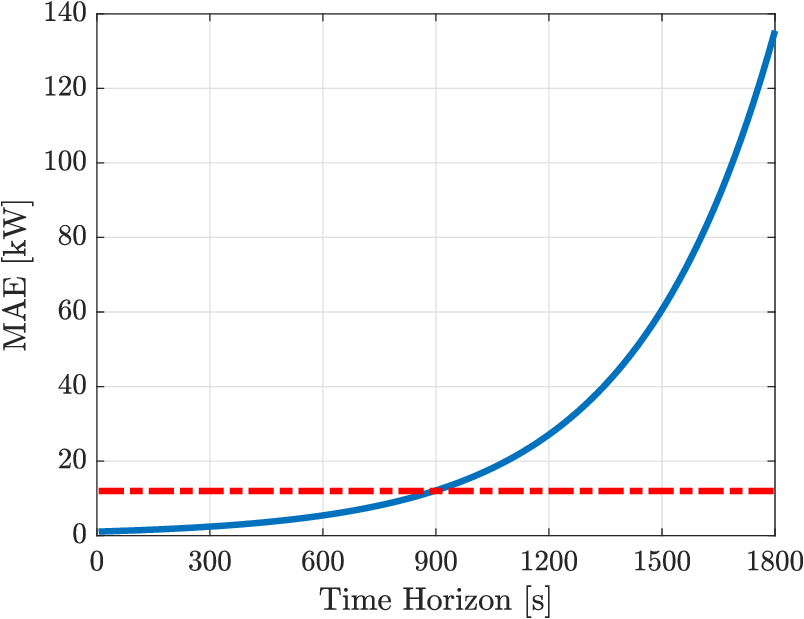}
\caption{MAE varying the time horizon for XGBoost+TSFM.}
\label{imm:exp:forecasting_mae}
\end{figure}
\begin{figure}
\centering
\includegraphics[width=.4\textwidth]{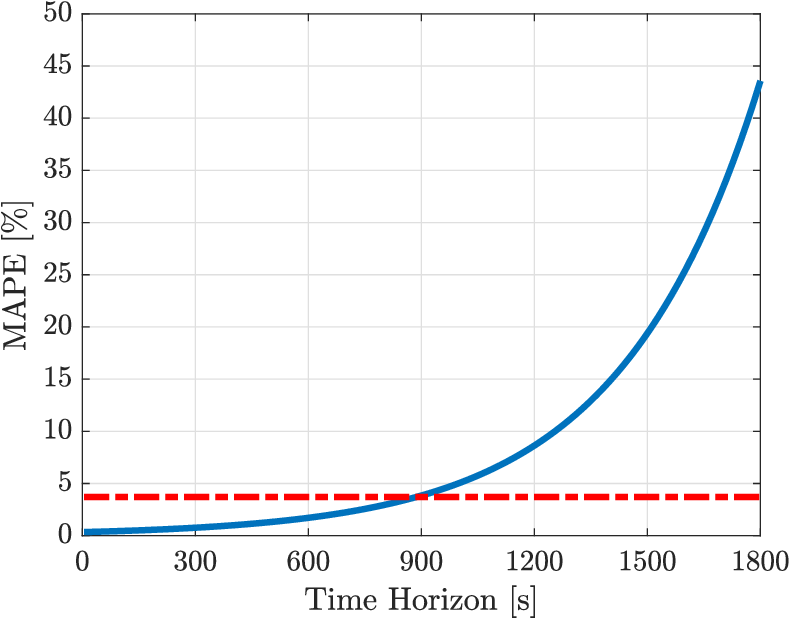}
\caption{MAPE varying the time horizon for XGBoost+TSFM.}
\label{imm:exp:forecasting_mape}
\end{figure}
In this section we will present the results of applying the method presented in Section~\ref{subsec:loadforecasting:methods} to the problem and data described in Section~\ref{subsec:loadforecasting:data}.
As a first step, we report in Table~\ref{tab:exp:forecasting:1} the MAE, MAPE, and PPMCC for XGBoost, TCN, TSFM, and the combination of XGBoost and TSFM for different time horizons.
It is further possible to observe that XGBoost is the best-performing method among the base methods (XGBoost, TCN, and TSFM).
Nevertheless, XGBoost+TSFM resulted in an improvement over the base methods.
From Table~\ref{tab:exp:forecasting:1}, we also note that on the \SI{30}{min} time horizon, the prediction starts to lose accuracy and therefore meaning.
The increase of MAE and MAPE for an increased time horizon with XGBoost+TSFM is shown in Figs.~\ref{imm:exp:forecasting_mae} and~\ref{imm:exp:forecasting_mape}, respectively.

\begin{figure*}
\centering
\begin{subfigure}{\textwidth}
\centering
\includegraphics[width=\textwidth]{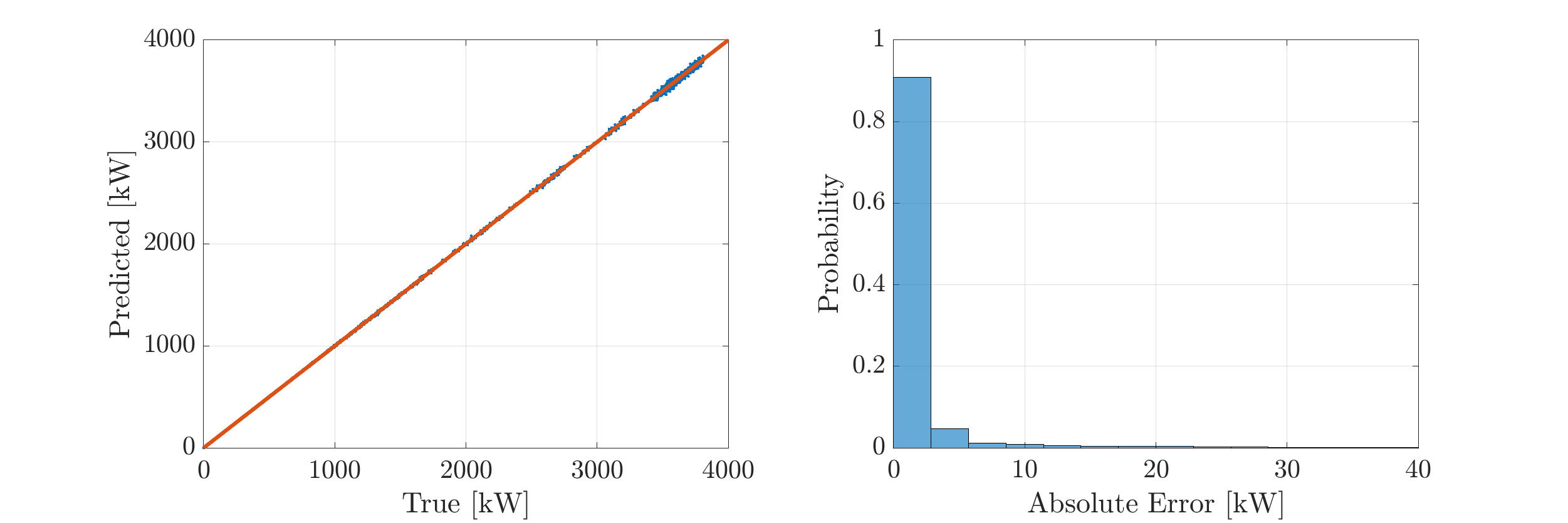}
\caption{5~s}
\label{imm:exp:forecasting:1.1}
\end{subfigure}\\
\begin{subfigure}{\textwidth}
\centering
\includegraphics[width=\textwidth]{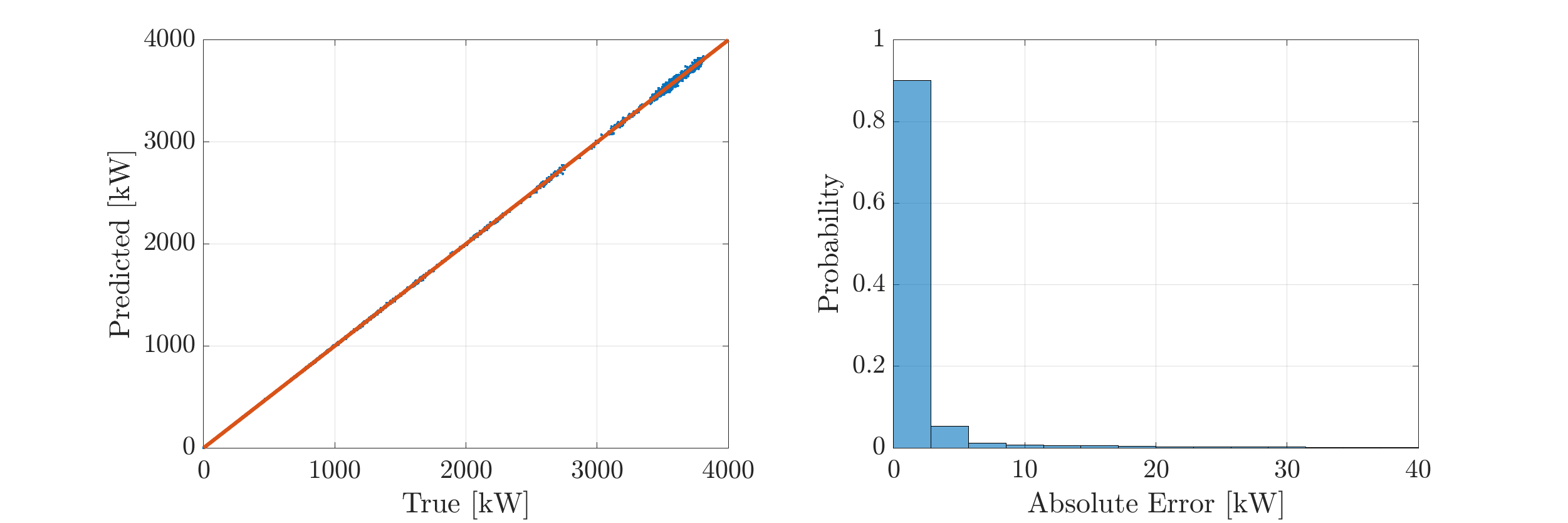}
\caption{1~min}
\label{imm:exp:forecasting:1.2}
\end{subfigure}\\
\begin{subfigure}{\textwidth}
\centering
\includegraphics[width=\textwidth]{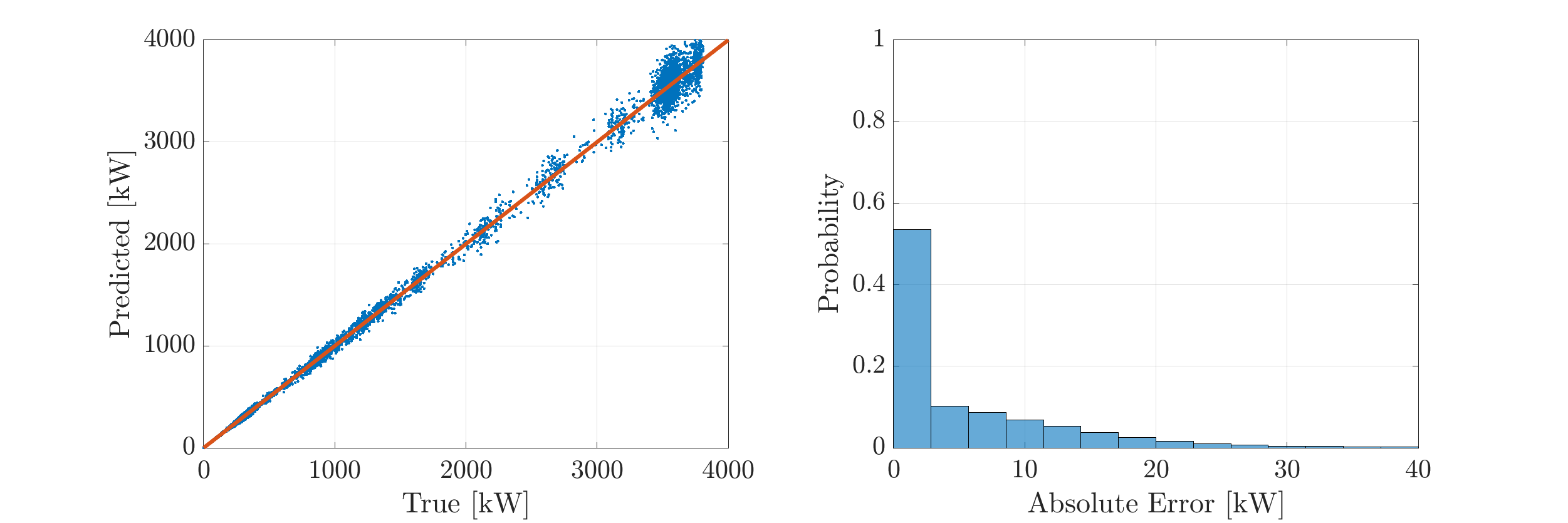}
\caption{15~min}
\label{imm:exp:forecasting:1.3}
\end{subfigure}
\caption{Load Forecasting: Scatter Plot and histogram of the absolute error for time horizons of a) \SI{5}{s} (a), b) \SI{1}{min}, and c) \SI{15}{min} for XGBoost+TSFM (best model according to Table~\ref{tab:exp:forecasting:1}).}
\label{imm:exp:forecasting:1}
\end{figure*}

\begin{figure*}
\centering
\begin{subfigure}{\textwidth}
\centering
\includegraphics[width=.45\textwidth]{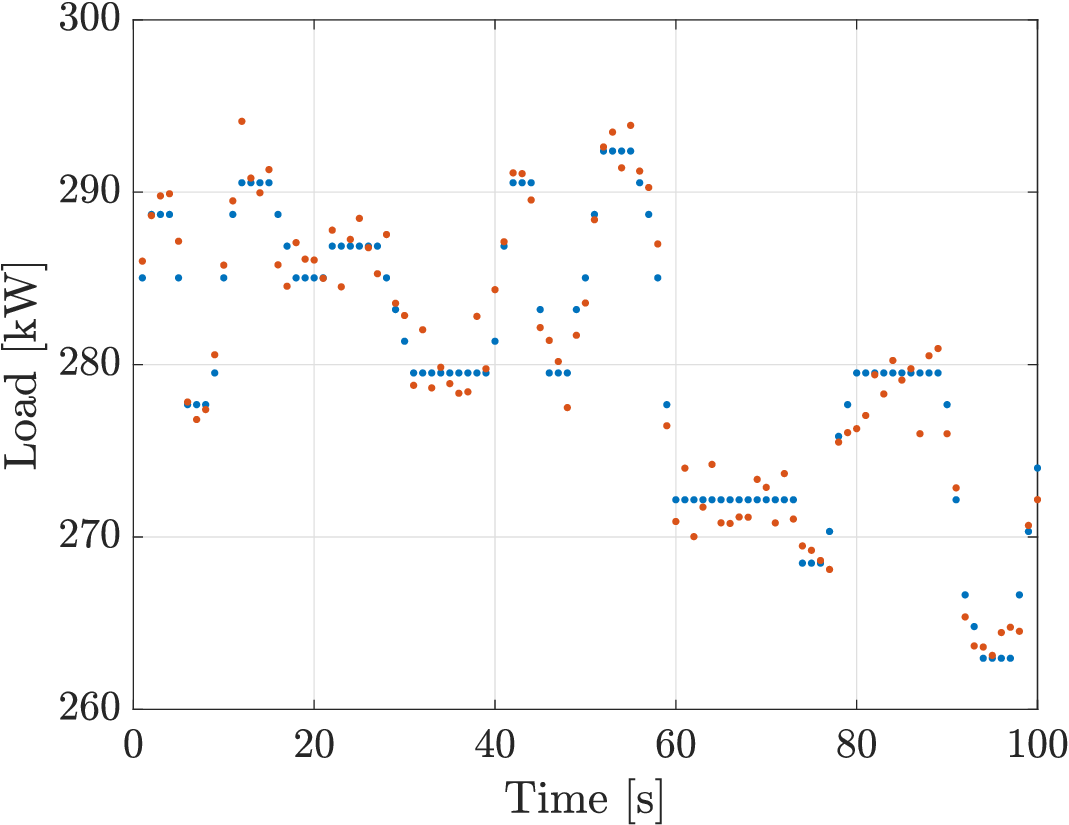}
\includegraphics[width=.45\textwidth]{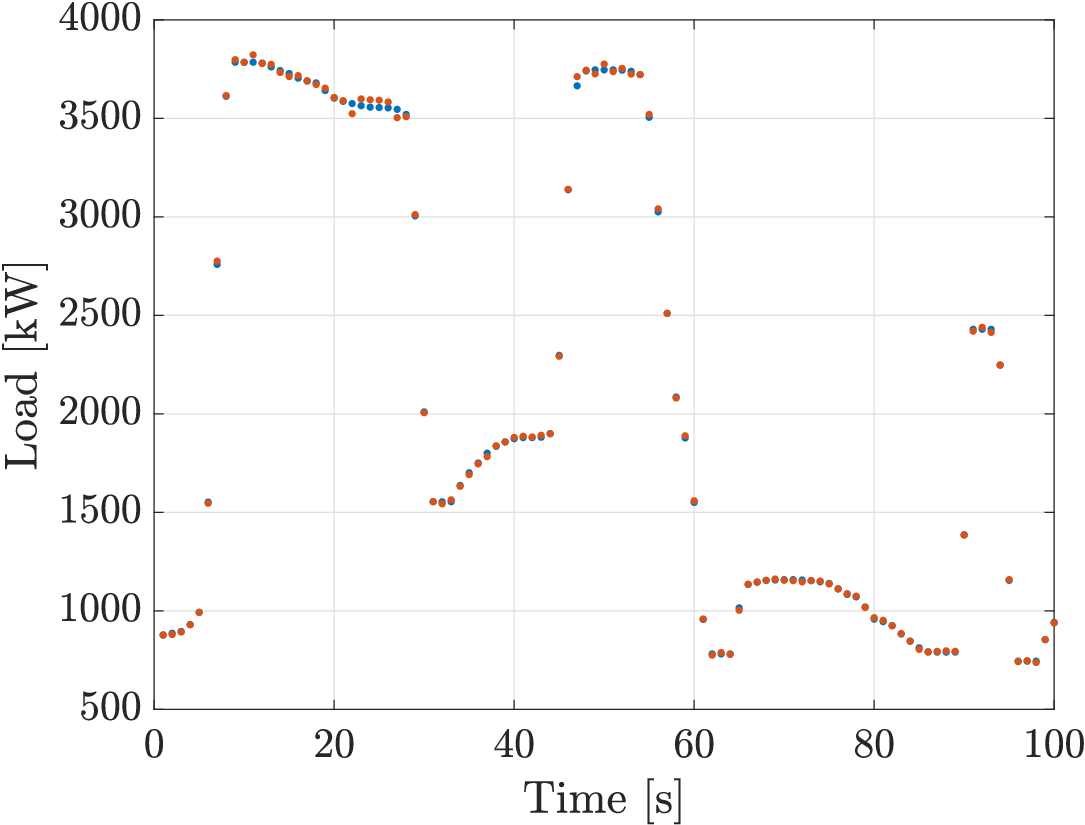}
\caption{5~s}
\label{imm:exp:forecasting:2.1}
\end{subfigure}\\
\begin{subfigure}{\textwidth}
\centering
\includegraphics[width=.45\textwidth]{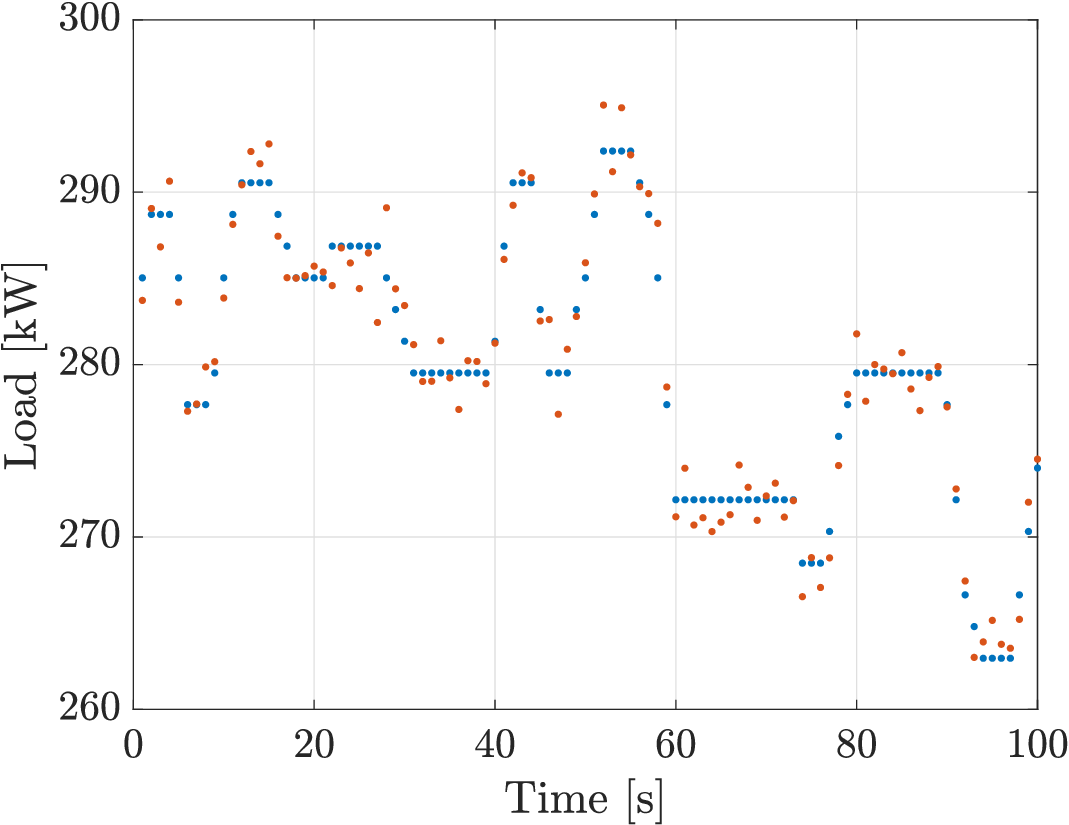}
\includegraphics[width=.45\textwidth]{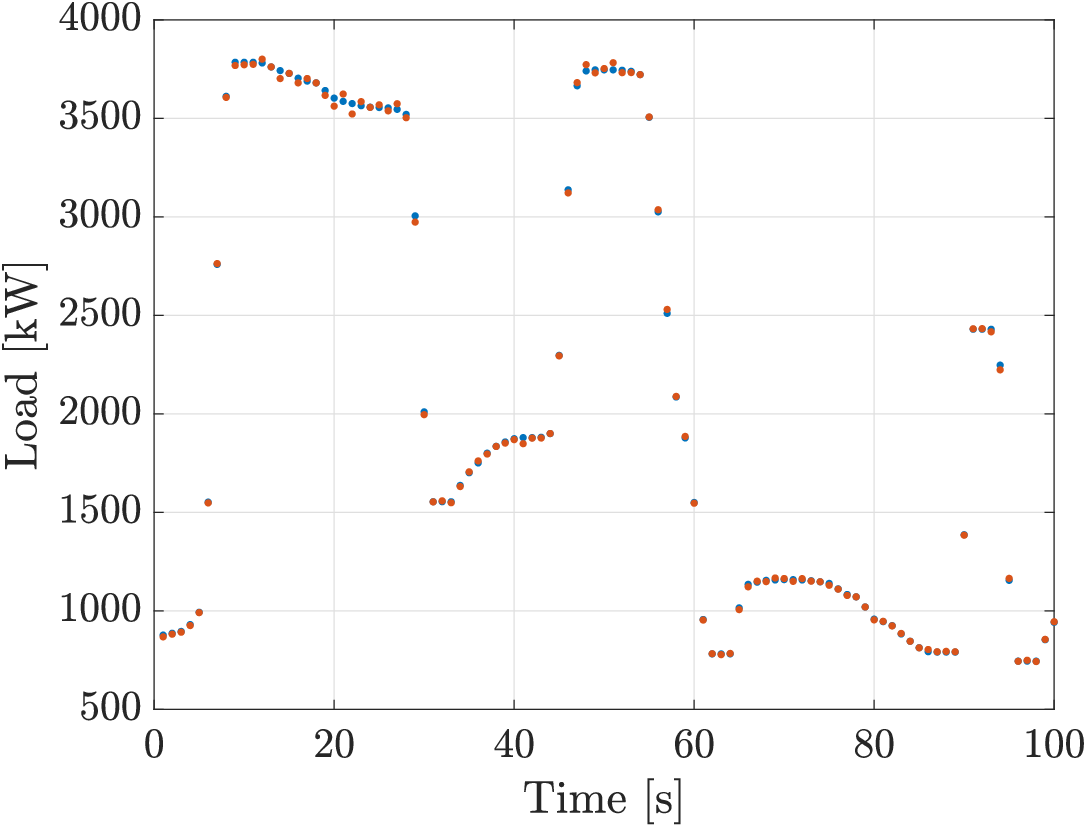}
\caption{1~min}
\label{imm:exp:forecasting:2.2}
\end{subfigure}\\
\begin{subfigure}{\textwidth}
\centering
\includegraphics[width=.45\textwidth]{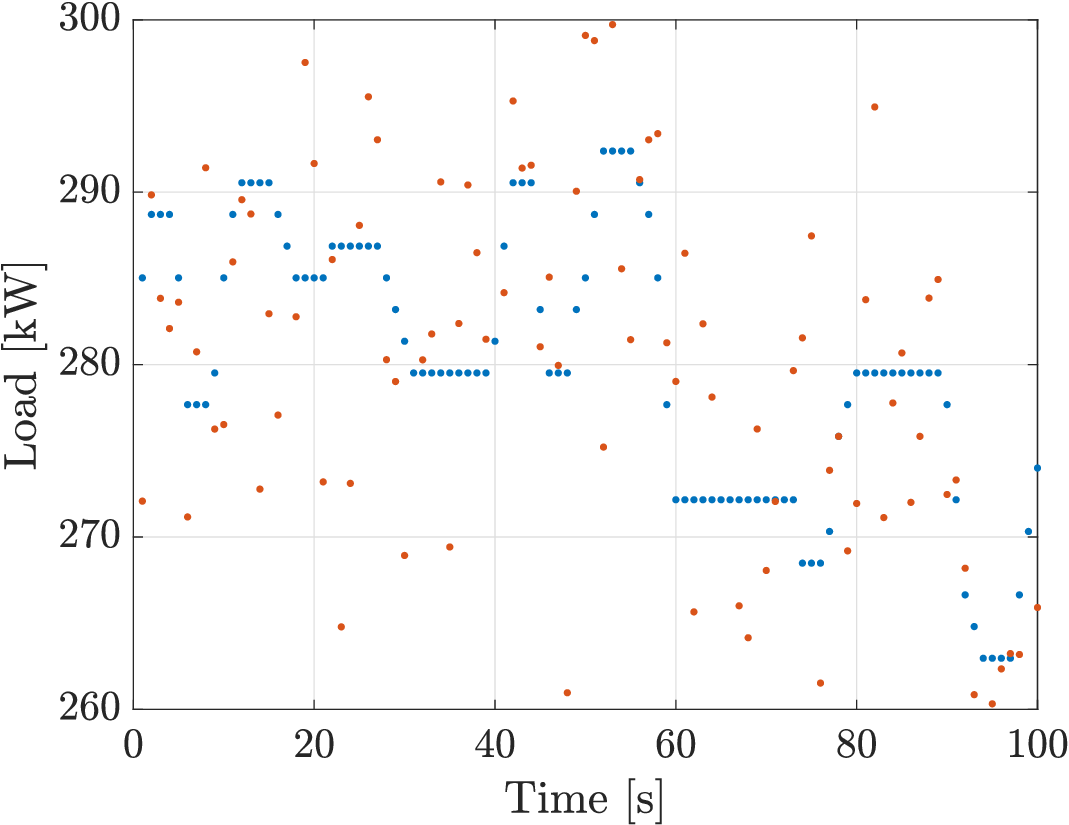}
\includegraphics[width=.45\textwidth]{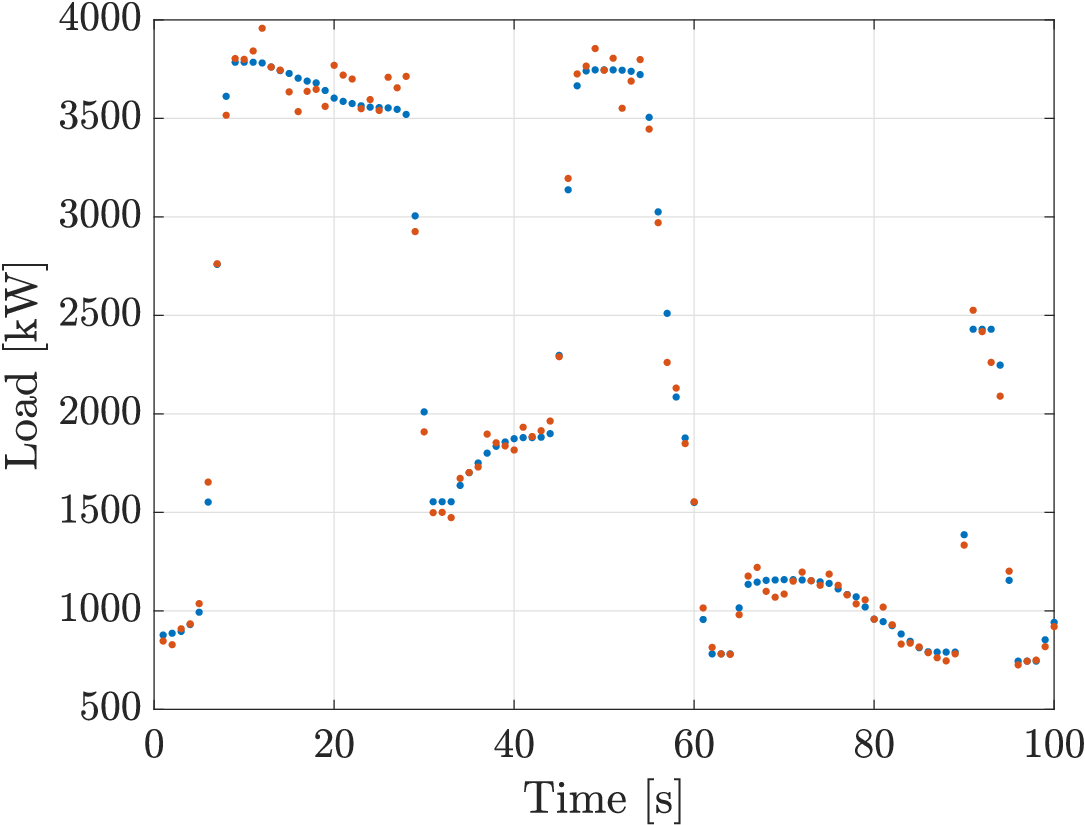}
\caption{15~min}
\label{imm:exp:forecasting:2.3}
\end{subfigure}
\caption{Load Forecasting: real load (blue) and predicted one (red) with XGBoost+TSFM prediction (the best model according to Table~\ref{tab:exp:forecasting:1}) for time horizons of a) \SI{5}{s} (a), b) \SI{1}{min}, and c) \SI{15}{min} at different regimes (left low load and right medium/high load).}
\label{imm:exp:forecasting:2}
\end{figure*}

To understand what time horizon can be considered meaningful, Figs.~\ref{imm:exp:forecasting_mae} and ~\ref{imm:exp:forecasting_mape} show the dependency of MAE and MAPE on the time horizon for XGBoost+TSFM (best model according to Table~\ref{tab:exp:forecasting:1}).
Figs.~\ref{imm:exp:forecasting_mae} and~\ref{imm:exp:forecasting_mape} indicate that \SI{15}{min} (\SI{900}{s}) is the limit to have a reasonable error (around \SI{10}{kW} and \SI{5}{\%}).
For the sake of completeness we report in Fig.~\ref{imm:exp:forecasting:1} scatter plots and Fig.~\ref{imm:exp:forecasting:2} the real versus predicted load for time horizons of \SI{5}{s}, \SI{1}{min} and \SI{15}{min} with XGBoost+TSFM prediction (best performing model according to Table~\ref{tab:exp:forecasting:1}).
From Figs.~\ref{imm:exp:forecasting:1} and~\ref{imm:exp:forecasting:2} it is possible to observe also qualitatively the very good agreement between the real values and the XGBoost+TSFM predictions on the considered time horizon.
\clearpage
\section{Power System Control}
\label{sec:power-sys-ctrl}
\begin{figure*}
	\centering
	\includegraphics{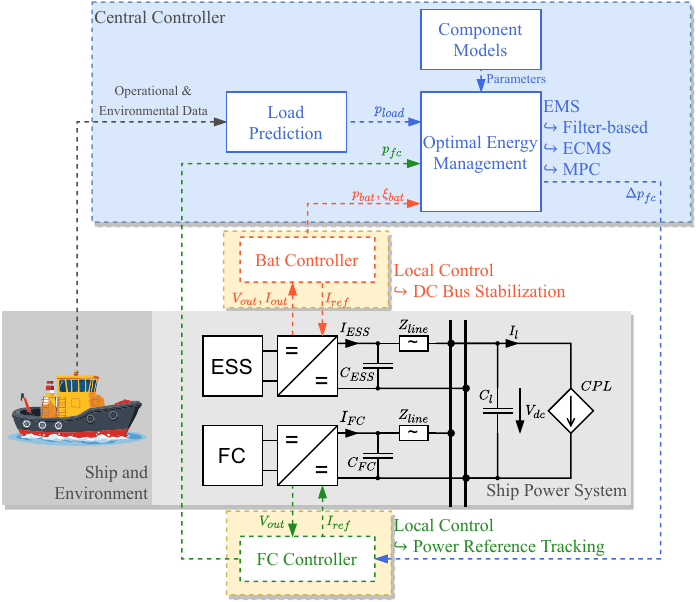}
	\caption{Depiction of the plant model and the control architecture for filter-based EMS, ECMS, and MPC including load predictions}
	\label{fig:control-architecture}
\end{figure*}

This chapter describes the energy management strategies which are used in this work to minimize the operating costs accounting for hydrogen consumption and cell degradation mechanisms.
The controlled system and centralized control architecture in this study is depicted in Fig.~\ref{fig:control-architecture}.
First, a filter-based controller is introduced against which the optimization-based strategies are benchmarked.
Next, a minimization problem accounting for the steady-state operating costs and the state-space model from chapter~\ref{sec:modeling} is constructed.
Finally, \gls{ECMS}-based approach is expanded to the predictive \gls{EMS} which leverages the data-driven load forecasting from chapter~\ref{sec:load-forecast} and explicitly considers the dynamic costs and constraints of the system.

\subsection{Filter-based Control}
A virtual impedance-based controller for scalable \gls{FC}-battery systems with DC distribution was presented in~\citet{kopkaDecentralizedPowerSharing2023} and~(JMET(submitted)).
These works realized a filter-based power split between the power generation and energy storage modules, with additional control outputs for voltage restoration and \gls{SoC} management.
The filter-based power split and \gls{SoC} management functionalities are implemented here as a benchmark control strategy.
The \gls{FC}'s output power is determined based on the measured load power filtered through a low-pass filter such that
\begin{equation}
	p_{fc}=\frac{\tau_{fd}}{s+\tau_{fd}}p_{load} + f(\xi_{bat})
\end{equation}
where $\tau_{fd}$ is the tunable frequency decoupling time constant. Superimposed on the frequency decoupling is an \gls{SoC} management strategy as laid out in~\citet{kopkaDecentralizedPowerSharing2023}.

\begin{figure}
	\centering
	\includegraphics{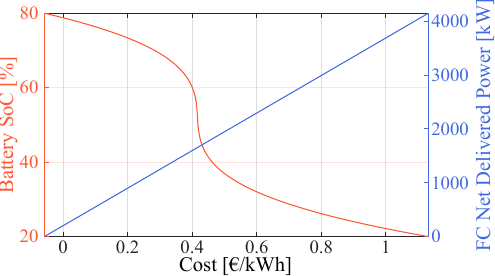}
	\caption{Equivalent cost of stored energy as a function of battery SoC versus marginal static cost of FC energy at BOL.}
	\label{fig:equivalence_factor}
\end{figure}

\subsection{ECMS}
To determine an instantaneously optimal split between the \gls{FC} and stored energy in the battery, a cost function needs to be defined.
Originally, the ECMS approach assigns an equivalent fuel consumption to the energy stored in the battery~\citet{sciarrettaControlHybridElectric2007}.
Here, we instead define an equivalent cost, which allows to incorporate both the hydrogen consumption as well as the FC degradation, thus making this approach health-aware.
The power-dependent cost of the \gls{FC} was already defined in Sections~\ref{sec:modeling-h2-consumption},~\ref{sec:modeling-fc-degradation}, and~\ref{sec:modeling-cost-model}.

The energy stored in the battery is assigned the equivalent cost $\lambda$.
The optimal power split at a given time instance and load power is then defined by the \gls{FC} power output minimizing the cost function
\begin{equation}
	\min_{\dot{p}_{fc}} f_{fc,stat}\left(p_{fc}\right) +\lambda \left(p_{bat}+\left(\frac{p_{bat}}{E_{oc}}\right)^2R_{bat,i}\right) + \mu \dot{p}_{fc}^2
	\label{eq:ecms-opt}
\end{equation}
accounting for the system dynamics~(\ref{eq:sys-dynamics}) and constraints (\ref{eq:Pfc_grad_lim}, \ref{eq:Pfc_lim}, \ref{eq:soc_lim}).
To avoid high \gls{FC} power gradients and unstable dynamic behavior of the controller, an arbitrary dynamic cost $\mu$ is added to the optimization.

\subsubsection{Equivalent Cost of Stored Energy}
A further challenge for the ECMS algorithm lies in the limited energy that can be stored in the batteries.
Accordingly, an additional \gls{SoC} management strategy is required to avoid the violation of charge limits.
This can be realized by adjusting the equivalent cost $\lambda$ based on the \gls{SoC}, such that \gls{ESS} power is expensive at low and cheap at high charge.

Neglecting the dynamic cost $\mu$ and battery losses, the optimal power split for~(\ref{eq:ecms-opt}) lies in the intersection of the marginal cost of the \gls{FC} $\frac{df_{fc,stat}(p_{fc})}{dp_{fc}}$ with $\lambda$.
Via an appropriate choice of $\lambda(\xi)$, we want to ensure that the \gls{FC} power is maximal at the minimum allows charge $\xi_{min}$, and zero at $\xi_{max}$.
Additionally, at a reference \gls{SoC}-level $\xi_{ref}$, the corresponding value of $\lambda(\xi)$ shall be equal to the marginal \gls{FC} cost at a reference power level $P_{fc,ref}$.
The function $\lambda(\xi)$ is therefore defined by the third-order polynomial displayed in Fig.~\ref{fig:equivalence_factor} where
\begin{equation}
	\lambda(\xi_{ref})=\frac{df_{fc,stat}(P_{fc,ref})}{dp_{fc}}
\end{equation}

\begin{equation}
	\lambda(\xi_{min})=\frac{df_{fc,stat}(P_{fc,max})}{dp_{fc}}
\end{equation}

\begin{equation}
	\lambda(\xi_{max})=\frac{df_{fc,stat}(0)}{dp_{fc}}
\end{equation}

\begin{equation}
	\frac{\lambda(\xi_{ref})}{d\xi}=0
\end{equation}

\subsubsection{Dynamic Costs}
The dynamic cost $\mu$ in~(\ref{eq:ecms-opt}) is intended to limit the magnitude of system inputs and avoid oscillations around an equilibrium point.
Further, it is an implicit means to incorporate the cost of dynamic load changes.
The formulation of the cost is similar to the dynamic cost in~\ref{eq:dynamic-deg}.
However, $\mu$ must be considerably smaller than $dv_{dyn2}$, otherwise the full cost of the load change would outweigh the static components in (\ref{eq:ecms-opt}), which only consider one time instant, yielding a sluggish controller.

\subsection{MPC}
Whereas the previous ECMS controller implements an instantaneous optimization, an MPC incorporates future inputs, states and disturbances as well.
This section describes the formulation of the MPC for the energy management of the FC-hybrid vessel.

\subsubsection{Optimization Problem}
Given a prediction horizon $N_{mpc}$, an optimization time period of $T_{mpc}$, the optimal input $\dot{p}_{fc}^*(k)$ at time instance $k$ solves the optimization problem

\begin{equation}
	\min_{\dot{p}_{fc}} \sum_{n=k}^{k+N_{mpc}-1}
	T_{mpc}\biggl((f_{fc,stat}(p_{fc}(n))+d\dot{V}_{dyn}(\dot{p}_{fc}(n))
	\shoveright{+ \lambda(\xi(k)) R_{i,bat} \left(\frac{p_{bat}(n)}{E_{oc}}\right)^2\biggr)}
	- \lambda(\xi(k))\xi(k+N_{mpc})E_{oc}C_{bat}
	\label{eq:mpc-opt}
\end{equation}
s.t.~(\ref{eq:Pfc_grad_lim}),~(\ref{eq:Pfc_lim}),~(\ref{eq:soc_lim}), and~(\ref{eq:sys-dynamics}).
For each time step $n$, the static and dynamic costs of the \gls{FC}, as well as an equivalent cost for battery losses are considered. Additionally, the equivalent value of stored energy at time $k+N_{mpc}$ is implemented as a terminal cost.
To avoid complex non-linearities, the equivalent cost $\lambda(\xi)$ is evaluated for the \gls{SoC} at time $k$ rather than for each time instance, such that it is eliminated as a dependent variable.
Doing so leaves only quadratic and linear terms in the cost function and linear constraints.
Accordingly, this problem can be solved using quadratic programming. The optimization is run once per optimization time interval $T_{mpc}$.

\section{Numerical Investigations}
\label{sec:num-investigations}
The described energy management strategies are implemented in a numerical simulation environment using MATLAB/Simulink. The settings for the simulation and the controllers are listed in Table~{\ref{tab:sim-params}}. Several scenarios are tested to highlight the functioning principles and the performance of the different strategies.
First, a rectangular load profile is applied to the system to highlight the working principles of the \gls{ECMS} and \gls{MPC}.
Next, results of an exemplary mission profile for all strategies are shown for demonstrating the behavior of the respective approaches in real conditions and subsequently, aggregated results over all available mission profile are described. Finally, the effect of a variation of the prediction and control horizon on the performance of the \gls{MPC} is analyzed to estimate the expected benefit of obtaining accurate predictions over a longer horizon.

\begin{table}
	\centering
	\caption{Simulation Settings}
	\begin{center}
		\begin{tabular}{ccc}
			\hline
			Parameter & Description & Value \\
			\hline
			$T_{sim}$ & Simulation step width & \SI{1}{s} \\
			$T_{ecms}$ & ECMS step width & \SI{5}{s} \\
			$T_{mpc}$ & MPC step width & \SI{30}{s} \\
			$N_{mpc}$ & Control horizon & \SI{30}{} \\
			$\mu$ & Dynamic cost in ECMS & $0.01\frac{dt_{dyn}dv_{dyn}}{P_{fc,max}^2}$ \\
			$\xi_{init}$ & Initial SoC & \SI{50}{\%} \\
			\hline
		\end{tabular}
		\label{tab:sim-params}
	\end{center}
\end{table}

\subsection{Rectangular Load Profile}
To highlight the functioning principles of the respective \gls{EMS} implementations, a simple rectangular-shaped load is fed into the system. For this, we emulate a mission sequence from sailing to towing to sailing. From the load histogram in Fig.~\ref{fig:data-histogram}, we derive a power of \SI{1300}{kW} during sailing and \SI{3800}{kW} during towing. We consider a shortened duration of towing of \SI{10}{min}, such that it falls within the prediction horizon and we assume a perfect power load prediction for this case.

\begin{figure}
	\centering
	\includegraphics{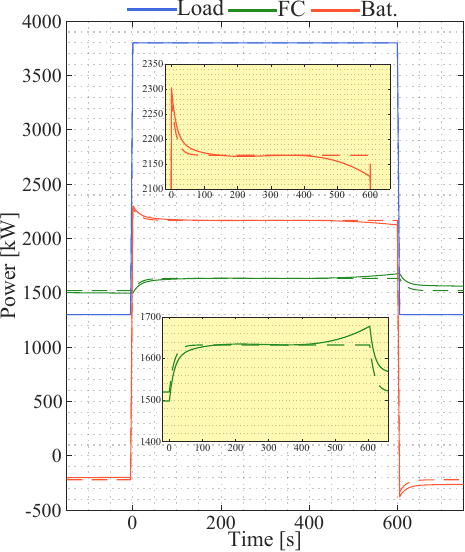}
	\caption{Power trajectories of FC and battery with ECMS for rectangular load shape with (solid) and without (dashed) SoC adaptation of equivalent cost. Zoom-ins (yellow-backed) on details of battery and FC trajectories.}
	\label{fig:rect_load_ecms_ink}
\end{figure}

\begin{figure}
	\centering
	\includegraphics{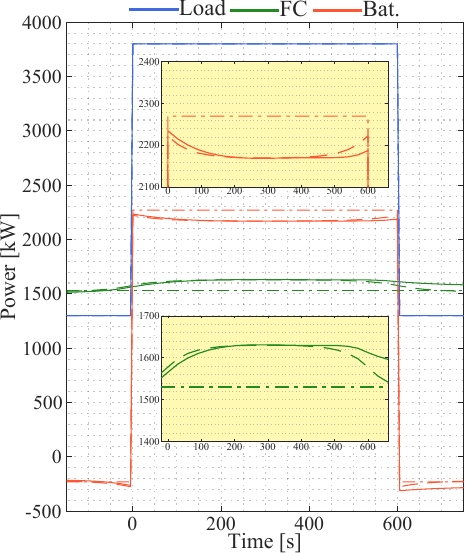}
	\caption{Power trajectories of FC and battery with MPC for rectangular load shape with cost adaptation (solid), without cost adaptation (dashed), and without cost adaptation plus disregard of battery losses (dash-dot). Zoom-ins (yellow-backed) on details of battery and FC trajectories.}
	\label{fig:rect_load_mpc_ink}
\end{figure}

The resulting power trajectories for the ECMS and \gls{MPC} with \SI{15}{min} prediction horizon are displayed in Figs.~\ref{fig:rect_load_ecms_ink} and~\ref{fig:rect_load_mpc_ink}, respectively. The plots show both the full \gls{EMS} implementations as well as a case in which the equivalent cost is not adapted to the battery \gls{SoC}. For the \gls{MPC}, an additional case is investigated in which the battery losses are ignored in the optimization problem.

With \gls{ECMS} it can clearly be seen that the controller is acting only upon a load change. At the load step, the batteries are immediately compensating the load imbalance, while the \gls{FC} increases its power to establish a cost balance. The speed of this adaptation is subject to the dynamic cost $\mu$. Without a cost factor adaptation, the system stays in an equilibrium as neither the \gls{FC} cost nor the battery cost change. With cost adaptation active, the \gls{FC} power is constantly increasing to compensate the increased cost of stored energy due to a decreasing \gls{SoC}. Over an increased amount of time, the \gls{FC} power would be required to match the load power, while the \gls{SoC} approaches its lower limit. After the negative load step it can be observed that the \gls{FC} power is higher than initially, due to the lower \gls{SoC}. Without \gls{SoC} adaptation, the trajectories are symmetrical.

Employing \gls{MPC}, the controllers act already before the load step occurs, anticipating the power imbalance. The trajectory smoothly approaches an equilibrium, limiting the power gradients of the \gls{FC}. As with the \gls{ECMS}, the cost factor adaptation leads to an asymmetrical power trajectory accounting for the decreased battery \gls{SoC}. Without cost adaptation and additionally neglecting battery losses, the \gls{FC} power is completely constant, since the prediction horizon is longer than the pulse. This showcases the fact that in the first two scenarios, the increase in \gls{FC} power output is done to avoid increased costs of battery losses.

\subsection{Mission Simulation}
\label{sec:res-mis-sim}
\begin{figure}
	\centering
	\includegraphics{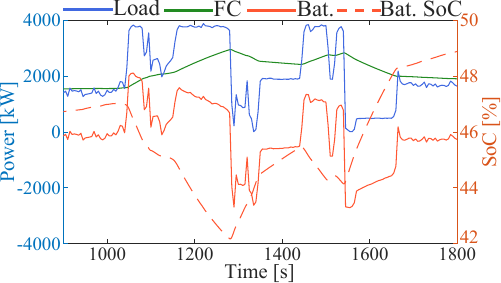}
	\caption{Power split between FC and battery, and battery SoC during high fluctuations with filter-based control and $\tau_{fd}=\SI{600}{s}$.}
	\label{fig:mission_fd600_zoom}
\end{figure}

\begin{figure}
	\centering
	\includegraphics{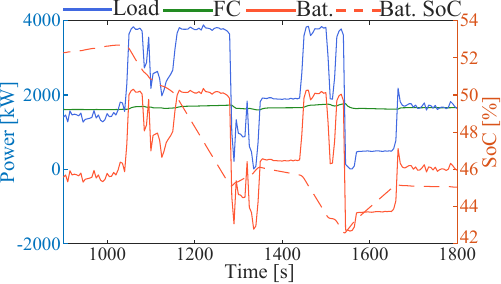}
	\caption{Power split between FC and battery, and battery SoC during high fluctuations with ECMS.}
	\label{fig:mission_ecms_zoom}
\end{figure}

\begin{figure}
	\centering
	\includegraphics{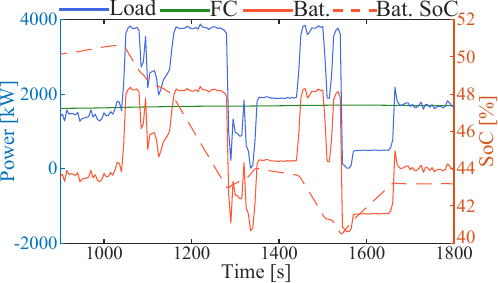}
	\caption{Power split between FC and battery, and battery SoC during high fluctuations with MPC and data-driven load prediction.}
	\label{fig:mission_mpc_data_zoom}
\end{figure}

\begin{figure*}
	\centering
	\includegraphics[width=\columnwidth]{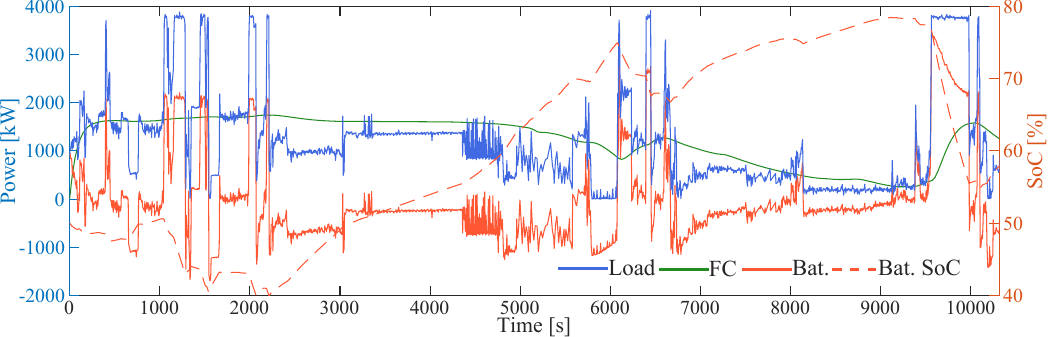}
	\caption{Power split between FC and battery, and battery SoC during mission simulation with MPC and data-driven load prediction.}
	\label{fig:mission_mpc_data}
\end{figure*}

The application of the strategies is demonstrated with one exemplary mission profile. Figs.~\ref{fig:mission_fd600_zoom},~\ref{fig:mission_ecms_zoom}, and~\ref{fig:mission_mpc_data_zoom} show the power trajectories during a mission phase with highly fluctuating loads. Fig.~\ref{fig:mission_mpc_data} shows the complete mission profile using \gls{MPC} leveraging data-driven load predictions.
The comparison shows that both optimization-based approaches yield a flattened \gls{FC} power output, with clearly reduced fluctuations and peak value.
The \gls{MPC} further minimizes fluctuations compared to the \gls{ECMS}-based controller such that fast fluctuations are almost entirely covered by the batteries.

\begin{figure}
	\centering
	\includegraphics{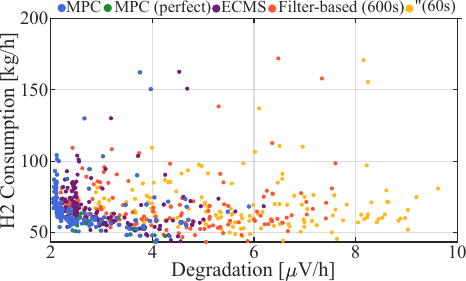}
	\caption{Hydrogen consumption rate versus fuel cell degradation rate for filter-based, ECMS, and MPC strategies for all mission profiles at BOL.}
	\label{fig:Pareto_Front_Scatter}
\end{figure}

\begin{figure}
	\centering
	\includegraphics{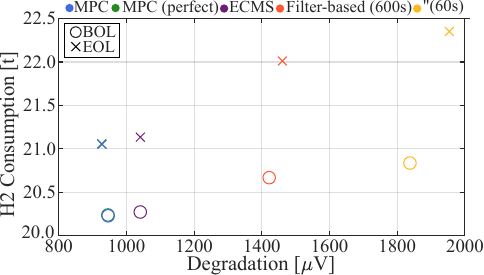}
	\caption{Cumulative hydrogen consumption versus fuel cell degradation for filter-based, ECMS, and MPC strategies over all mission profiles.}
	\label{fig:Pareto_Front_Total}
\end{figure}

To compare the performance of all strategies, the strategies are applied to all 167 available mission profiles, and their performance in terms of hydrogen consumption and \gls{FC} degradation are compared, for \gls{BOL} and \gls{EOL} conditions.
In total, the missions add up to a total operation time of \SI{345.1}{h}. This time only accounts for active operation of the vessel and covers only the available measurements during the period described in Section~\ref{subsec:loadforecasting:data}.
Fig.~\ref{fig:Pareto_Front_Scatter} shows a scatter plot indicating the results of all simulations at \gls{BOL} for all described strategies, i.e. frequency decoupling with time constans of $\tau_{fd}=\SI{60}{s}$, $\tau_{fd}=\SI{600}{s}$, \gls{ECMS}, and \gls{MPC} with perfect and with data-driven load predictions. Fig.~\ref{fig:Pareto_Front_Total} shows the accumulated hydrogen consumption and degradation over all missions for these strategies.
The comparison of simulation outcomes clearly shows the decrease in both cell degradation and hydrogen consumption that is achieved by the proposed strategies.
At \gls{BOL}, the benchmark controller with $\tau_{fd}=\SI{600}{s}$ yields \SI{1422}{\mathrm{\mu} V} cell degradation and \SI{20.67}{t} hydrogen consumption, which is already a clear improvement over the faster filter-based controller (\SI{1838}{\mathrm{\mu} V};\SI{20.84}{t}).
However, the ECMS (\SI{1041}{\mathrm{\mu} V};\SI{20.27}{t}) reduces these values by \SI{26.8}{\%} and \SI{1.9}{\%}, respectively.
The \gls{MPC} implementations with perfect prediction (\SI{946}{\mathrm{\mu} V};\SI{20.23}{t}) and data-driven prediction (\SI{945}{\mathrm{\mu} V};\SI{20.24}{t}) yield nearly identical results which are \SI{33.5}{\%} and \SI{2.1}{\%} under the benchmark.
At \gls{EOL}, it can be seen that the hydrogen consumption is generally higher. Whereas the filter-based approaches also show an increased degradation, the optimization-based approaches manage to limit or even reduce the latter.
The \gls{MPC}, compared to the benchmark, reduces total degradation by \SI{36.4}{\%}from \SI{1460}{\mathrm{\mu} V} to \SI{928}{\mathrm{\mu} V} and hydrogen consumption by \SI{5.8}{\%} from \SI{22.35}{t} to \SI{21.05}{t}.
The results show that the proposed \glspl{EMS} have high potential for improving the system performance, especially for the \gls{FC} durability.
Furthermore, the improvements become more significant when the \gls{FC}'s degradation is already in an advanced stage.

\subsection{Variation of Prediction Horizon}
\begin{figure}
	\centering
	\includegraphics{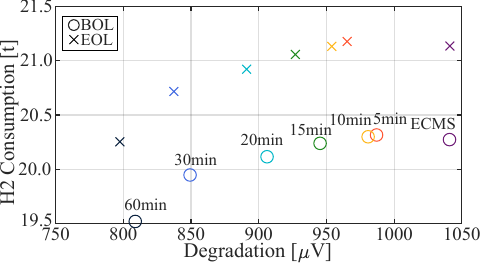}
	\caption{Cumulative hydrogen consumption versus fuel cell degradation for MPC with perfect prediction for varying prediction horizon.}
	\label{fig:Pareto_Front_Horizon}
\end{figure}

As evaluated in Section~\ref{sec:res-mis-sim}, the results of \gls{MPC} with data-driven load predictions are almost identical to those of \gls{MPC} with assumed perfect predictions. 
This underlines the applicability of data-driven load predictions in real-time applications. 
In a next step, the effect of the prediction horizon on the control performance is analyzed. 
Since data-driven predictions are currently only available over \SI{15}{min}, a comparison with a horizon of up to \SI{60}{min} is conducted assuming perfect predictions. 
The results at \gls{BOL} and \gls{EOL} are shown in Fig.~\ref{fig:Pareto_Front_Horizon}, indicating a significant boost in with an increased prediction horizon. 
At \gls{BOL} with \SI{15}{min} prediction horizon, a cell degradation of \SI{945}{\mathrm{\mu} V} and \SI{20.24}{t} hydrogen consumption can be realized. 
An increased prediction horizon can decrease the degradation by up to \SI{14.4}{\%} to \SI{809}{\mathrm{\mu} V} and fuel use up to \SI{3.6}{\%} to \SI{19.52}{t} with a \SI{1}{h} horizon. 
At \gls{EOL} the possible improvements are in a similar range with \SI{14.0}{\%} and \SI{3.8}{\%}, respectively.
The graph shows that especially the degradation can be significantly lowered by including longer forecasting periods. 
A prediction over \SI{15}{min} has already been proven to be realistic in Section~\ref{sec:load-forecast}, while an increased horizon significantly increases the prediction error (Fig.~\ref{imm:exp:forecasting_mape}). 
This comparison shows the expected benefits arising from an increased accuracy over longer horizons, e.g., due to more data or additional insights.

\section{Conclusions}
\label{sec:conclusions}
To increase the economic feasibility of \gls{FC}-battery hybrid \gls{SPS}, it is paramount to decrease its operating costs. A key factor in this is the design of the \gls{EMS}, in which not only the hydrogen consumption but also the cell degradation, as a major contributor to the operating costs, must be considered.
Leveraging data-driven load forecasting, this work implemented a degradation-aware predictive energy management that aims at minimizing the total operating costs of a zero-emission harbor tug.
Using a degradation-aware equivalent cost approach for the instantaneous optimization of the energy management, a simultaneous decrease of both hydrogen consumption and cell degradation could be achieved compared to a filter-based benchmark.
The data-driven load prediction, given the available data time-series of the tugboat, showed a fairly accurate load prediction with a MAPE of \SI{2.8}{\%} at \SI{15}{min}.
Leveraging this prediction, the degradation-aware predictive \gls{EMS} yielded additional improvements in both fuel consumption as well as cell degradation over the \gls{ECMS}-based method.
It was further shown that an accurate load prediction over even longer predictions horizons of up to an hour could yield additional, significant performance gains.
At an advanced aging state of the \gls{FC} system towards its \gls{EOL}, the relative difference in performance between the tested strategies increased, underscoring the potential benefits from the predictive strategy.
By utilizing real measurement data and limiting the inputs to those available in the virtually retrofitted zero-emission tugboat, this work proved the real-time applicability of the predictive energy management in actual systems. This is essential to realize the expected reductions in fuel consumption and cell degradation, thereby lowering the operational costs and contributing to the cost-competitiveness of zero-emission ships.

For future work it is intended to expand this approach to further ship types. Whereas this study assumes a central control of a single \gls{FC} and battery, larger system require the coordination of multiple parallel generation and storage devices. This will require to account for different component ratings and capabilities, potentially needing alternative control architectures, e.g. in a distributed architecture. Furthermore, this work expanded the equivalent cost approach of the \gls{ECMS} to its \gls{MPC} framework, including a terminal cost for the battery \gls{SoC}. As a next step, it will be interesting to determine whether a multi-horizon approach with an outer mission-scale optimization will allow for an improved performance by computing a target \gls{SoC} and on-off state of \gls{FC} systems for the inner \gls{MPC}.
Whereas it seems that the minimization of both fuel consumption and cell degradation is favoring similar control strategies, i.e. steady operation within an efficient operation band, the inclusion of on-off switching of \gls{FC} modules is expected to require a clearer trade-off between objectives.
Furthermore, this work did not explicitly assign costs to the efficiency decrease due to degradation, but this warrants further exploration. In addition, the fixed \SI{10}{\%} performance drop for \gls{EOL} definition of \glspl{FC} is to be questioned for this application and instead be replaced by an optimization variable.

\section*{Acknowledgement(s)}
This research is supported by the \textit{Sustainable Hydrogen Integrated Propulsion Drives (SH2IPDRIVE)} project, which has received funding from RvO (reference number MOB21013), through the RDM regulation of the Ministry of Economic Affairs and Climate Policy.

\section*{Declaration of competing interest}
The authors declare that they have no known competing financial interests or personal relationships that could have appeared to
influence the work reported in this paper.


\bibliographystyle{cas-model2-names}

\bibliography{ZoteroLibrary,DCMicrogridDistributed,biblio}


%
%
%

\end{document}